\documentclass[english]{article}
\usepackage[T1]{fontenc}
\usepackage{amsmath}
\usepackage[latin9]{inputenc}
\usepackage{geometry}
\usepackage{float}
\usepackage{calc}
\usepackage{mathrsfs}
\usepackage{amsmath}
\usepackage{amssymb}
\usepackage{graphicx}
\usepackage{array}
\usepackage{caption}

\makeatletter

\providecommand{\tabularnewline}{\\}

\usepackage{ dsfont }
\usepackage[all]{xy}
\usepackage{cite}
\usepackage{colonequals}

\usepackage{textgreek}

\usepackage{tikz}

\makeatletter
\let\@fnsymbol\@arabic
\makeatother

\RequirePackage{mathtools}

\RequirePackage{mathrsfs}
\RequirePackage[scr=boondoxo,scrscaled=1.05]{mathalfa}

\AtBeginDocument{

}

\makeatother

\usepackage{babel}
\begin{document}
\global\long\def\u{u}
\global\long\def\p{P}
\global\long\def\X{X}
\global\long\def\me{\mu_{e}}
\global\long\def\sym{\textrm{sym}}
\global\long\def\grad{\nabla}
\global\long\def\le{\lambda_{e}}
\global\long\def\tr{\textrm{tr}}
\global\long\def\mc{\mu_{c}}
\global\long\def\skew{\textrm{skew}}
\global\long\def\curl{\textrm{Curl}}
\global\long\def\ac{\alpha_{c}}
\global\long\def\B{\mathscr{B}}
\global\long\def\R{\mathbb{R}}
\global\long\def\fr{\rightarrow}
\global\long\def\Q{\mathcal{Q}}
\global\long\def\A{\mathscr{A}}
\global\long\def\L{\mathscr{L}}
\global\long\def\D{\mathscr{D}}
\global\long\def\id{\mathds{1}}
\global\long\def\ds{\textrm{dev sym}}
\global\long\def\sph{\textrm{sph}}
\global\long\def\eg{\boldsymbol{\varepsilon}}
\global\long\def\aa{\boldsymbol{\alpha}}
\global\long\def\o{\boldsymbol{\omega}}
\global\long\def\ege{\boldsymbol{\varepsilon}_{e}}
\global\long\def\egp{\boldsymbol{\varepsilon}_{p}}
\global\long\def\punto{\,.\,}
\global\long\def\Sym{\textrm{Sym}\left(3\right)}
\global\long\def\MM{\boldsymbol{\mathfrak{M}}}
\global\long\def\C{\mathbb{C}}
\global\long\def\gl{\mathfrak{gl}\left(3\right)}
\global\long\def\P{P}
\global\long\def\Lin{\textrm{Lin}}
\global\long\def\D{\boldsymbol{D}}
\global\long\def\a{\alpha}
\global\long\def\b{\beta}
\global\long\def\lle{\lambda_{e}}
\global\long\def\ce{\mathbb{C}_{e}}
\global\long\def\cm{\mathbb{C}_{\textrm{micro}}}
\global\long\def\cc{\mathbb{C}_{c}}
\global\long\def\axl{\textrm{axl}}
\global\long\def\dev{\textrm{dev}}
\global\long\def\mh{\mu_{\textrm{micro}}}
\global\long\def\lh{\lambda_{\textrm{micro}}}
\global\long\def\vau{\omega_{1}^{int}}
\global\long\def\vad{\omega_{2}^{int}}
\global\long\def\axl{\textrm{axl}}
\global\long\def\Le{\mathbb{L}_{e}}
\global\long\def\Lc{\mathbb{L}_{c}}
\global\long\def\V{\mathbb{V}}
\global\long\def\as{\dagger}
\global\long\def\dd{\:\overset{{\scriptscriptstyle \triangledown}}{{\scriptscriptstyle \vartriangle}}\,}
\global\long\def\cM{\mathbb{C}_{\textrm{macro}}}
\global\long\def\mum{\mu_{\textrm{macro}}}
\global\long\def\lam{\lambda_{\textrm{macro}}}

\newcommand{\Csecond}{\widetilde{\mathbb{C}}}
\newcommand{\Jsecond}{\widetilde{\mathbb{J}}}
\newcommand{\Dsecond}{\widetilde{\mathbb{D}}}
\newcommand{\Cfourth}{\overline{\mathbb{C}}}
\newcommand{\Csym}{\mathbb{C}}
\newcommand{\Cso}{\Csym}
\newcommand{\Jfourth}{\overline{\Jsym}}
\newcommand{\Jsym}{\mathbb{J}}
\newcommand{\Jso}{\Jsym}
\newcommand{\Lf}{\overline{\mathbb{L}}}
\newcommand{\Ls}{\widehat{\mathbb{L}}}
\newcommand{\Lsym}{\mathbb{L}}
\newcommand{\E}{\overline{\mathbb{E}}_{\mathrm{cross}}}

\newcommand{\Cc}{\Cso_{\mathrm{c}}}
\newcommand{\Ctc}{\Csecond_{c}}
\newcommand{\Ce}{\Csym_{\mathrm{e}}}
\newcommand{\Cte}{\Csecond_{\mathrm{e}}}
\newcommand{\Ltc}{\widetilde{\Lsym}_{c}}
\newcommand{\Lte}{\widetilde{\Lsym}_{e}}
\newcommand{\Coe}{\Cfourth_{e}}
\newcommand{\Ch}{\Csym_{\mathrm{micro}}}
\newcommand{\Cth}{\Csecond_{\mathrm{m}}}
\newcommand{\Ct}{\Csecond_{\mathrm{M}}}
\newcommand{\TensSym}{\widetilde{\mathbb{E}}}
\newcommand{\TensSkew}{\widetilde{\mathbb{K}}}
\newcommand{\Jh}{\Jsym_{\mathrm{m}}}
\newcommand{\Jth}{\Jsecond_{\mathrm{m}}}
\newcommand{\Jg}{\Jso_{\mathrm{g}}}
\newcommand{\Jtg}{\Jsecond_{\mathrm{g}}}
\newcommand{\Je}{\Jsym_{\mathrm{e}}}
\newcommand{\Jte}{\Jsecond_{\mathrm{e}}}
\newcommand{\Jc}{\Jso_{\mathrm{c}}}
\newcommand{\Jtc}{\Jsecond_{\mathrm{c}}}
\newcommand{\mm}{\mu_{\mathrm{M}}}
\newcommand{\lm}{\lambda_{\mathrm{M}}}
\newcommand{\ke}{\kappa_{e}}
\newcommand{\kh}{\kappa_{\mathrm{m}}}
\newcommand{\km}{\kappa_{\mathrm{M}}}
\newcommand{\mLc}{\mu\, L_{c}^{2}}
\newcommand{\mLd}{\mu\, L_{d}^{2}}
\newcommand{\sig}{\widetilde{\sigma}}
\newcommand{\Sig}{\sigma_{\mathrm{M}}}
\newcommand{\n}{ n}

\newcommand{\nablau}{\,\nabla u}
\newcommand{\nablap}{\nabla P}
\newcommand{\Curl}{\,\mathrm{Curl}}
\newcommand{\Div}{\mathrm{Div}}
\newcommand{\devsym}{\dev\sym}
\renewcommand{\skew}{\, \mathrm{skew}}
\newcommand{\x}{\cdot}
\newcommand{\langlenew}{\,\big\langle\,}
\newcommand{\ranglenew}{\,\big\rangle}

\newcommand{\so}{\mathfrak{so}}
\renewcommand{\sl}{\mathfrak{sl}}

\title{Relaxed micromorphic model of transient wave propagation in anisotropic band-gap metastructures}
\author{Gabriele Barbagallo\thanks{Gabriele Barbagallo, gabriele.barbagallo@insa-lyon.fr, GEOMAS, INSA-Lyon, Universit$\rm \acute{e}$ de Lyon, 20 avenue Albert Einstein,
		69621, Villeurbanne cedex, France}~, Domenico Tallarico\thanks{Domenico Tallarico, corresponding author, domenico.tallarico@insa-lyon.fr, GEOMAS, INSA-Lyon, Universit$\rm \acute{e}$ de Lyon, 20 avenue Albert Einstein,
		69621, Villeurbanne cedex, France}~, Marco Valerio d\textquoteright Agostino\thanks{Marco Valerio d'Agostino,  marco-valerio.dagostino@insa-lyon.fr,
		GEOMAS, INSA-Lyon, Universit$\rm \acute{e}$ de Lyon, 20 avenue Albert Einstein, 		69621, Villeurbanne cedex, France}~, Alexios Aivaliotis\thanks{Alexios Aivaliotis, alexios.aivaliotis@insa-lyon.fr, GEOMAS, INSA-Lyon, Universit$\rm \acute{e}$ de Lyon, 20 avenue Albert Einstein,
		69621, Villeurbanne cedex, France},\\Patrizio Neff \thanks{Patrizio Neff, patrizio.neff@uni-due.de, Head of Chair for Nonlinear Analysis and Modelling, Fakult{\"a}t f{\"u}r Mathematik, Universit{\"a}t Duisburg-Essen,
Mathematik-Carr{\' e}e, Thea-Leymann-Stra{\ss}e 9, 45127 Essen, Germany}$\;$$\;$and Angela Madeo\thanks{Angela Madeo, angela.madeo@insa-lyon.fr, GEOMAS, INSA-Lyon, Universit$\rm \acute{e}$
de Lyon, 20 avenue Albert Einstein, 69621, Villeurbanne cedex, France}}
\maketitle
\begin{abstract}
In this paper, we show that the transient waveforms arising from several localised pulses in a micro-structured material can be reproduced by a corresponding generalised continuum of the relaxed micromorphic type. Specifically, we compare the dynamic response of a bounded micro-structured material to that of bounded continua with special kinematic properties: (i) the relaxed micromorphic continuum and (ii) an equivalent Cauchy linear elastic continuum. We show that, while the Cauchy theory is able to describe the overall behaviour of the metastructure only at low frequencies, the relaxed micromorphic model goes far beyond by giving a correct description of the pulse propagation in the frequency band-gap and at frequencies intersecting the optical branches. In addition, we observe a computational time reduction associated with the use of the relaxed micromorphic continuum, compared to the sensible computational time needed to perform a transient computation in a micro-structured domain.
\end{abstract}
\addtocounter{footnote}{5} \vspace{6mm}

\vspace{2mm}
\textbf{}
\newpage{}
\tableofcontents

\section{Introduction\label{sec:intro}}

\addtocounter{footnote}{-5} 

Nowadays, metamaterials hold a central role in engineering developments thanks to their exotic mechanical and electromagnetic behaviour. For instance, the phenomena experimentally observed include: dynamic anisotropy, negative Poisson ratio, focusing of mechanical or electromagnetic waves via negative refraction, active/passive cloaking and many more. One of the most appealing mechanical features is the possibility to filter waves in specific ranges of frequencies, referred to as frequency ``stop-bands'' or ``band-gaps''. Exploiting such unorthodox properties, the conception of morphologically complex finite structures in many fields of engineering could be revolutionised. For instance, a finite ordered set of a metamaterial's unit-cell, that throughout this paper we will call ``metastructure'', would inherit the filtering properties of the periodic counterpart, leading to application ranging from seismic protection to stealth.

Forced (\emph{i.e.} non-homogeneous) scalar (out-of-plane) and vector (in-plane) problems of linear elasticity in micro-structured unbounded domains have been studied extensively in the literature. 
The Green's function provides a formidable analytical tool to write down an integral representation of the solution which can be evaluated, via analytical or numerical tools, in different frequency regimes. Colquitt \emph{et al.} in \cite{colquitt2012dynamic} obtained the out-of-plane  displacement field  in square and triangular elastic lattices, with a particular emphasis on the resonances associated with dynamic anisotropy and \emph{primitive waveforms} arising at saddle points in the dispersion diagram. These primitive waveforms have also been examined in the papers by Langley~\cite{langley1996response}, Ruzzene \emph{et al.}~\cite{ruzzene2003wavebeaming}, Ayzenberg-Stepanenko and Slepyan~\cite{ayzenbergstepanenko2008resonant}, and Osharovich \emph{et al.}~\cite{osharovich2010II}, among others. We also mention the papers by Martin~\cite{martin2006discrete} and Movchan and Slepyan~\cite{movchan2009band}, which analyse the properties of the vector dynamic Green's functions for a square lattice in the pass and stop band, respectively. Nonetheless, the use of such analytical tools for continua with a complex microstructure can be challenging, especially in the transient regime. Therefore, transient and/or time-harmonic finite element solutions for such problems can be used.

One of the most widespread tools used to study the ability of periodic media to support or inhibit free wave propagation is the  Bloch-Floquet analysis \cite{brillouin2003wave}. This technique allows to reduce the free wave propagation problem, posed in a periodic domain, to an eigenvalue problem in its period, referred to as unit-cell, with quasi-periodic boundary conditions at the boundaries of the unit-cell.

In recent articles, it was shown that the dynamic response of a band-gap metamaterial can be described as a continuous medium using the relaxed micromorphic model. The relaxed micromorphic model was introduced and described in \cite{neff2014unifying,neff2015relaxed,neff2017real,ghiba2014relaxed,barbagallo2017transparent}. Furthermore, for the well posedness of the relaxed micromorphic model and for the mathematical framework needed to describe frequency band-gaps in metamaterials, we refer the reader to \cite{madeo2017modeling,madeo2017role,madeo2016reflection,madeo2016first,madeo2016complete}. None of the other generalized continuum models (as second gradient models or Mindlin-Eringen model) have been able to well describe band-gaps in metamaterials to the same extent of the relaxed micromorphic model. Indeed, second gradient models cannot grasp complex dispersion due to their simplified kinematics, allowing only acoustic branches to be accounted for. On the other hand, the Mindlin-Eringen model, although it considers an appropriate kinematical framework, it relies on a large number of constitutive parameters, thus making \color{black}{very tricky any physical} \color{black} interpretation. Due to its simplified constitutive relations for the strain energy density, the relaxed micromorphic model allows to account for the main overall dispersion properties of band-gap metastructures, via a relatively contained number of elastic parameters. Such parameters simplification relies on the ``sym-skew-trace''  orthogonal decomposition adopted for the micro and macro deformation measures of the relaxed micromorphic model.

As its main original contribution, the present paper shows that the relaxed micromorphic model is rich enough to face exciting problems concerning the mechanical behaviour of two-dimensional (2D) anisotropic micro-structured materials, which possess frequency band-gaps and non-trivial dispersive properties. More specifically, we compare the high-frequency dynamic response of a specific micro-structured material - subjected to space-concentrated and time-modulated loads - to the corresponding response issued by the relaxed micromorphic continuum. In particular, it is shown that the transient waveforms within the chosen metastructure are well-captured by the relaxed micromorphic continuum model, while the classical Cauchy continuum model is only accurate at lower frequencies. 

\color{black}{
As already stated, the present work relies on preceding results which were obtained by some of the authors in previous papers. Such results are summarized in Tab. \ref{tab:papers}.\\
\begin{table}
\begin{center}
    \begin{tabular}{ | l |  p{6cm}  | p{6cm} | }
    \hline
    \textbf{Paper} & \textbf{Main results} & \textbf{Novelty} \\ \hline
   Present paper & The tetragonal identification of the Relaxed Micromorphic Model (RMM) parameters performed in \cite{dagostino2018effective}, is used here to model transient waveforms in the RMM continuum, which are in qualitative agreement with transient waveforms in a micro-structured material. & First evidence that the identification of material parameters of the RMM on real 2D microstructures can ultimately lead to an effective continuum PDEs description of the transient behavior of complex metastructures, with significant computational time saving.\\ \hline
 \cite{dagostino2018effective}& Application of the results provided in \cite{barbagallo2017transparent} for general anisotropy to the tetragonal symmetry class. Application to a real tetragonal metamaterial  & Effective determination of the RMM using informations from a real microstructure with tetragonal symmetry.\\ \hline
   \cite{barbagallo2017transparent} & Rigorous establishment of the anisotropic framework for the RMM and application to the tetragonal case.  & Introduction of arbitrary anisotropy in the RMM thus opening the possibility of applying the RMM to metamaterials structures belonging to any class of symmetry.\\
    \hline
  \cite{madeo2017modeling,madeo2017role,madeo2016complete}  &Subsequent studies investigating the importance of the constitutive form of kinetic energy of the RMM for the effective description of real systems.& Proof of the fact that gradient micro-inertia is of primary importance to correctly fit the behavior of real metamaterials for the complete range of wavelengths spanning from infinity to the size of the unit cell.\\ \hline
  \cite{madeo2016first,madeo2016reflection}&Study of diffractive properties of RMM material surfaces in the isotropic case.& Proof that the RMM provided with suitable BCs is able to describe complex phenomena of reflection and transmission in metamaterials.\\ 
  \hline
 \cite{ghiba2014relaxed,neff2014unifying}& Proof of mathematical well-posedness of the RMM.& The RMM is a rather recent model and a proof that its solution is unique - provided that adequate boundary conditions are used - is of pivotal importance for a correct application in mechanics.\\
 \hline 
      \end{tabular}
      \caption{\label{tab:papers} The table highlights the novelty and main achievement so far of the relaxed micromorphic model.}
      \end{center}
      \end{table}
      We want to stress here the fact that, when a new theory is being constructed, partial superposition with previous works is unavoidable. It is exactly such incessant progress that will ultimately allow to proceed towards the ambitious result of establishing a generalized version of the theory of elasticity for the study of microstructured metamaterials. In fact, as the theory of elasticity allows the averaged description of the mechanical behavior of classical materials without accounting for the complexity of all the atoms and molecules that constitute them, our long term goal is to build up a generalized theory of elasticity enclosing the dynamic response of metamaterials but without accounting for the detailed microstructure. Given the ambition of this final goal, it is clear that a series of papers is needed, with the aim to set up the model, to encompass its generalization to general anisotropy, to establish a simplified fitting procedure on real metamaterials, and to providing evidence that the obtained material parameters identification is able to reproduce real behaviors at large scales. Further research will be needed to prove the effectiveness of the established procedures to other metamaterials with different classes of symmetry and also independent tests of the effectiveness of the model (for example studying the reflective properties of interfaces embedded in anisotropic metamaterials) will be needed to valorize the results obtained up to now.}
 \color{black}
\begin{figure}[h!] 
\centering
\includegraphics[width=0.8\textwidth]{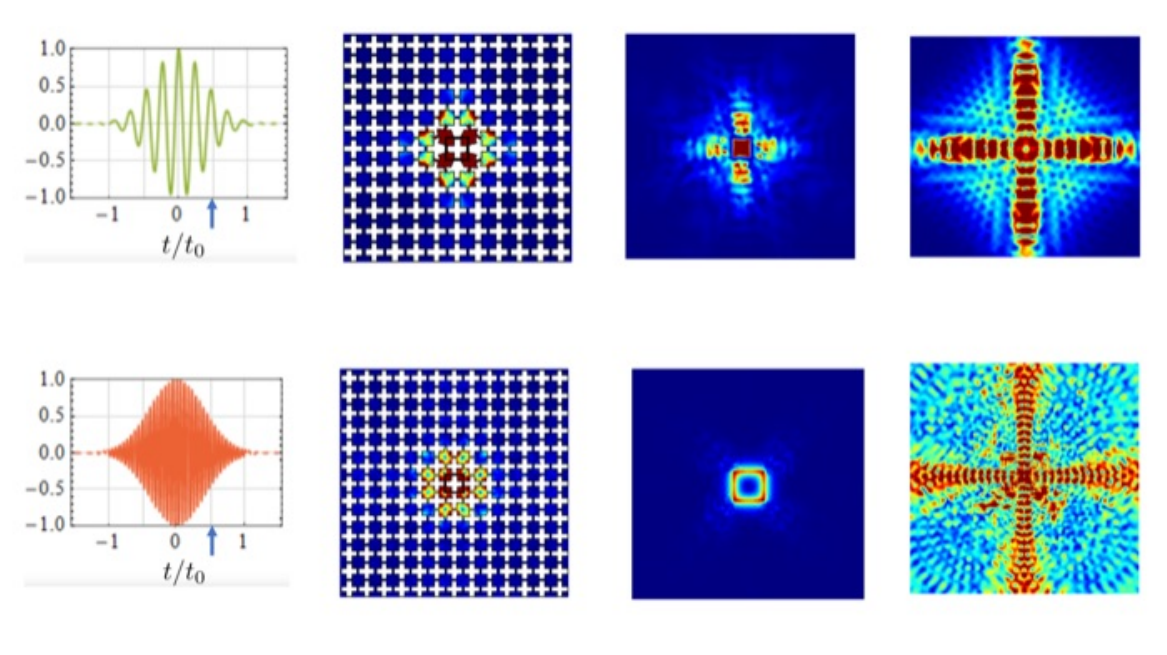}
	\begin{picture}(0,0)(0,0)
	\put(-350,120) {$(a)$}
	\put(-250,120) {$(b)$}
	\put(-155,120) {$(c)$}
	\put(-60,120) {$(d)$}
	\put(-350,5) {$(e)$}
	\put(-250,5) {$(f)$}
	\put(-155,5) {$(g)$}
	\put(-60,5) {$(h)$}
	\end{picture}
	\caption{\label{fig:pulse-intro} High-frequency waveforms in the transient regime for the considered tetragonal metastructure (panels (b)  and (f)) agree with those obtained in a relaxed micromorphic continuum model (panels (c)  and (g)). The aforementioned waveforms are generated by the application of space-concentrated pulses, whose time-dependence is shown in panels (a) and (e), respectively. The blue arrows in panels (a) and (e) indicate the time at which the waveforms on their right have been evaluated. Panels (d) and (h) report the transient waveform in an equivalent Cauchy continuum, showing propagation instead of localisation. We refer to section \ref{sec:results} for a detailed discussion.}
\end{figure}
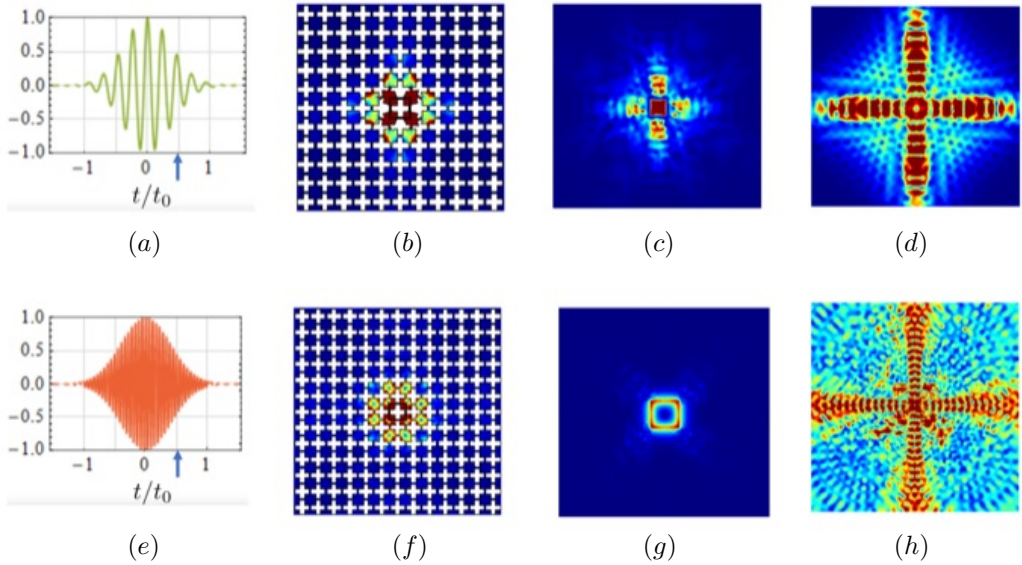

An outline of the manuscript is given as follows. In section \ref{sec:aniso-rm}, we define the action functional of the relaxed micromorphic continuum, define the constitutive fourth-order elastic tensors which appear therein, and give the system of partial differential equations (PDEs) resulting from the minimisation of the action functional. We particularise our analysis to 2D plain-strain continua belonging to the tetragonal symmetry class. Both an effective low-frequency Cauchy system and the relaxed micromorphic system of PDEs are given. In section \ref{sec:bf}, we introduce the micro-structured material of interest, define its unit cell and perform Bloch-Floquet analysis. The dispersive properties of the periodic structure are shown, via dispersion curves along specified paths in the first Brillouin zone. In addition, we compare the Bloch-Floquet dispersion diagram with the dispersion diagram issued via the relaxed micromorphic continuum model, enabling us to determine the constitutive parameters of our enriched model by a simple inverse approach. 
In section \ref{sec:pulse-def}, we introduce a space-concentrated and time-modulated pulse to be applied to (i) a finite metastructure, (ii) a finite continuum governed by the relaxed micromorphic model and (iii) a finite continuum modeled by an equivalent Cauchy model. The time-dependence of the prescribed pulse is tailored in order to explore different frequency regions within the dispersion diagram of the corresponding periodic structure (see Fig. 1(a) and 1(e), for example). Section \ref{sec:results} is devoted to the discussion of the main results of this article: we show that a relaxed micromorphic continuum subjected to the aforementioned transient pulse, displays similar waveforms to those exhibited by the micro-structured domain, including localised waveforms corresponding to frequencies in the band-gap and optical branches. On the other hand, the Cauchy model is not descriptive outside the low frequency regime. An illustrative anticipation of the obtained results is shown in Fig. 1. In addition, we show that by using our enriched continuum model instead of the explicit calculation of all the details of the microstructure provides a true advantage in terms of computational time, especially when considering large metastructures. Finally, in section \ref{sec:conclusions} we provide our main conclusions and an outline for future research directions.

\section{The anisotropic relaxed micromorphic model \label{sec:aniso-rm}}

The relaxed micromorphic model is defined in terms of  \textit{symmetric elastic (relative) strains} $\varepsilon_{e}:=\sym \left(\nablau-\p\right)$, where $u$ is the displacement and $P$ is  the $\mathbb{R}^{3\times3}$ non-symmetric micro-distortion, so that standard 4$^{th}$ order symmetric elasticity tensors can be used in order to define elastic stresses. Moreover, regarding the curvature, it considers the \textit{second order dislocation-density tensor} $\alpha=-\Curl \p$ instead of the third order curvature tensor $\nabla \p$ (used in the full Mindlin-Eringen model) with the effect that the description of the anisotropy of curvature only needs 4$^{th}$ order tensors, instead of 6$^{th}$ order ones. The potential energy is 
\begin{align}
W= & \underbrace{\frac{1}{2}\langlenew\Ce \, \sym\left(\nablau-\p\right),\sym\left(\nablau-\p\right)\ranglenew_{\R^{3\times3}}}_{\mathrm{{\textstyle elastic-energy}}}+\frac{1}{2}\underbrace{\langlenew\Ch\, \sym\, \p,\sym\, \p\ranglenew_{\R^{3\times3}}}_{\mathrm{\textstyle micro-self-energy}}\label{eq:EnerRelaxed}\\
& \ 
+\underbrace{\frac{1}{2}\langlenew\Cc  \skew\left(\nablau-\p\right),\skew\left(\nablau-\p\right)\ranglenew_{\R^{3\times3}}}_{
	\mathrm{\textstyle local\ rotational\ elastic\ coupling } } 
+\underbrace{\frac{\mLc}{2} \langlenew \, \Curl\, \p,\Curl\, \p\ranglenew_{\R^{3\times3}}}_{\mathrm{\textstyle curvature}}\,.
\nonumber 
\end{align}
Here  $\Ce,$ $\Ch:~\Sym\rightarrow\Sym$ are both classical $4^{th}$ order elasticity tensors \textit{acting on symmetric second order tensors} only: $\Ce$ acts on the \textit{symmetric elastic strain} $\varepsilon_{e}:=\sym \left(\nablau-\p\right)$ and $\Ch$ acts on the \textit{symmetric micro-strain} $\sym\, \p$ and both map to symmetric tensors. The tensor  $\Cc:\so(3)\rightarrow\so(3)$ is a $4^{th}$ order tensor that acts only on skew-symmetric matrices and yields only skew-symmetric tensors.
\color{black}{Moreover,  the parameter $L_c$ with dimensions of length can be interpreted as a characteristic length-scale which accounts for non-local effects in the considered metamaterial. Since, in this paper, we will only deal with bulk propagation phenomena, the effect of such length-scale is expected to be negligible. We leave to further publications the task of showing that the parameter $L_c$ may have a non-negligible role when scattering phenomena at the interface between homogeneous and microstructured materials are considered. Hence, in the reminder of this paper we will set $L_c=0$.}\color{black}~In Eq. \eqref{eq:EnerRelaxed} we have introduced $\langle\cdot,\cdot\rangle$ to denote the scalar product between tensor fields.

The dynamical formulation of the proposed relaxed micromorphic model is obtained introducing an anisotropic kinetic energy density of the type:
\begin{align}
J=&\hspace{-0.3cm}\underbrace{\frac{1}{2}\rho\left\Vert u_{,t}\right\Vert ^{2}}_{\mathrm{\textstyle{Cauchy\ inertia}}}\hspace{-0.4cm}+\hspace{0.1cm}\underbrace{\frac{1}{2}\langlenew \Jh\, \sym P_{,t},\sym P_{,t} \ranglenew+\frac{1}{2}\langlenew \Jc\, \skew P_{,t},\skew P_{,t} \ranglenew}_{\mathrm{\textstyle{free\ micro-inertia}}}\hspace{0.1cm}\label{eq:KineticAniso}
\\\nonumber
&+\hspace{0.1cm}\underbrace{\frac{1}{2}\langlenew \Je\, \sym \nablau_{,t}, \sym \nablau_{,t}\ranglenew+\frac{1}{2}\langlenew \Jg\, \skew \nablau_{,t},\skew \nablau_{,t}\ranglenew}_{\mathrm{\textstyle{gradient\ micro-inertia}}}.
\end{align}
Here  $\Je,$ $\Jh:~\Sym\rightarrow\Sym$ are both $4^{th}$ order micro-inertia tensors \textit{acting on symmetric second order tensors}, while the tensors  $\Jc$, $\Jg:\so(3)\rightarrow\so(3)$ are $4^{th}$ order tensors that act only on skew-symmetric matrices and yield only skew-symmetric tensors. In Eq. \eqref{eq:KineticAniso} we have introduced the notation $\left\Vert\cdot\right\Vert^2=\langle \cdot,\cdot\rangle$.

Given a bounded domain $\Omega$ with boundary $\partial \Omega$ and considering the anisotropic  strain energy \eqref{eq:EnerRelaxed} and kinetic energy \eqref{eq:KineticAniso}, the equations of motion for the relaxed micromorphic model are obtained by requiring that the action functional 
\begin{equation}\label{eq:action}
{\cal A}=\int_{0}^T {\rm d}t\int (J-W){\rm d}\Omega,
\end{equation}
is stationary with respect to the arbitrary variations of the kinematic fields. In Eq. \eqref{eq:action}, $[0,T]$ is the time interval during which the dynamics of the system is considered. In their strong form, the equations of motion  are 
\begin{figure}[H]
	\centering{}%
	\noindent\fbox{\begin{minipage}[t]{1\columnwidth - 2\fboxsep - 2\fboxrule}%
			\vspace{-0.5cm}
			\begin{align}\label{eq:PDE system}
			\rho\,u_{,tt}-\Div\left[\mathcal{I}\right] & =\Div\left[\,\sigma\,\right], & \Jh\,\sym   \p_{,tt}+\Jc\skew   \p_{,tt} & =\sigma-s-\Curl\,m,
			\end{align}
	\end{minipage}}
\end{figure}
with the associated natural and kinematic boundary conditions:
\begin{align}\label{eq:BCs}
f&:=\big(\,\sigma+\mathcal{I}\big)\x \n =t^{\mathrm{ext}}(x,t)\hspace{-1cm} &\mathrm{or} \qquad\quad&\,u=\varphi(x,t),& \quad &\forall x\in\partial\Omega_0, \nonumber\\
\tau\cdot\nu_i&:= -m\x\epsilon\x\n= \tau^{\mathrm{ext}}(x,t) \hspace{-1cm} &\mathrm{or}  \qquad\quad&\p\x \nu_{i}=p_{i}(x,t), \qquad \qquad i=2,3, & \quad &\forall x\in\partial\Omega_0,
\end{align}
where $\epsilon$ is the \color{black}{Levi-Civita permutation symbol}\color{black}, $n$ is the normal to the boundary, $\nu_1$ and $\nu_2$ are 2 orthogonal vectors tangent to the boundary $\partial \Omega_0$ (a contour where the external perturbation is applied). In addition, ${\varphi},p_2,p_3,t^{\mathrm{ext}},\tau^{\mathrm{ext}}$ are assigned quantities. In Eq. \eqref{eq:PDE system}  we have introduced
\begin{align}
\sigma & =\Ce \, \sym\left(\nablau-\p\right)+\Cc \skew\left(\nablau-\p\right),&
s & =\Ch\, \sym \p ,\\\mathcal{I}&=\Je\,\sym  \nablau_{,tt}+\Jg\skew  \nablau_{,tt},&
m & =\mLc \, \Curl \p.\nonumber 
\end{align}
\subsection{An equivalent macroscopic Cauchy model}
\color{black} We explicitly recall that, when considering wavelengths which are larger than the typical microstructure size, the influence of the microstructure itself  become minimal and the relaxed micromorphic model is equivalent to a classical Cauchy continuum with elastic stiffness tensor $\mathbb{C}_{\rm macro}$ (see \cite{barbagallo2017transparent}). A clear identification of the equivalent macroscopic fourth order tensor $\C_{\textrm{macro}}$ via the constitutive tensors $\Ch$ and $\Ce$ can also be derived as a harmonic tensor mean, \emph{i.e.}
\begin{align}
\C_{\textrm{macro}}  :=\Ch \left(\Ch+\Ce\right)^{-1} \, \Ce\,. \label{eq:Relation}	
\end{align} 
The homogenisation formula \eqref{eq:Relation} can be used to characterise the material with classical experimental (or numerical) procedures and to obtain a useful relation between the parameters of the relaxed micromorphic model and those of the equivalent Cauchy continuum which can be considered representative of the dynamic response of the metastructure at low frequencies. The equations of motion for the macroscopic Cauchy model is given by the PDE system:
\begin{align}\label{eq:PDEs-cmacro}
\rho\,u_{,tt}&=\textrm{Div}\left[\mathbb{C}_{\textrm{macro}}\,\sym\nabla u\right]& \forall x&\in\Omega,\\\nonumber
f:&=\left(\mathbb{C}_{\textrm{macro}}\,\sym\nabla u\right)\cdot n=t^{\rm ext}(x,t) \qquad \mathrm{or} \qquad u=\varphi(x,t),& \forall x&\in\partial\Omega_0,
\end{align}
where the same notation  as in Eq. \eqref{eq:BCs} for the external fields assigned to $\partial \Omega_0$ is used.

\subsection{The plane strain Ansatz for a relaxed micromorphic continuum with tetragonal symmetry \label{sec:plane-strain}}

In the remainder of this paper, we will focus on the in-plane behavior of a metamaterial introducing the following constraints:
\begin{itemize}
\item we consider only the in-plane degrees of freedom, i.e. $u_3,\ P_{13},\ P_{31},\ P_{23},\ P_{32}$ and $ P_{33}$ are set to zero;
\item we consider only the in-plane deformations i.e. all the derivatives on the out-of-plane direction are set to be zero.
\end{itemize}
These assumptions correspond to the  following simplified form of the kinematical fields:
\begin{equation}\label{eq:plane-strain}
u(x_1,x_2,t)=\begin{pmatrix}
u_1(x_1,x_2,t)\\u_2(x_1,x_2,t)\\0
\end{pmatrix},\qquad\qquad
P(x_1,x_2,t)=\begin{pmatrix}
P_{11}(x_1,x_2,t)&P_{12}(x_1,x_2,t)&0\\
P_{21}(x_1,x_2,t)&P_{22}(x_1,x_2,t)&0\\
0&0&0
\end{pmatrix}.
\end{equation}
Moreover, since we are interested in studying the dynamic response of the metastructure generated by the unit cell given in Fig. \ref{fig:Microstructure}, we consider that the corresponding relaxed micromorphic continuum must be particularised to the same symmetry class of the unit cell, \emph{i.e.} the tetragonal symmetry. 

With the proposed assumptions, the constitutive tensors of the relaxed micromorphic model  in the tetragonal symmetry case are given by\footnote{The ``tilded'' matrices are the Voigt representations of the corresponding forth order elastic tensors which appear in Eqs  \eqref{eq:EnerRelaxed} and \eqref{eq:KineticAniso}, defining the potential and kinetic energy, respectively. For a detailed discussion about passing from classical fourth order elastic tensor to the corresponding second order Voigt representation see \cite{barbagallo2017transparent}. The star symbols denote the components which do not intervene in the 2D plane-strain case.}:
\begin{align*}
\Cth & =\begin{pmatrix}2\,\mu_{\textrm{micro}}+\lambda_{\textrm{micro}} & \lambda_{\textrm{micro}} & \star & 0 & 0 & 0\\
\lambda_{\textrm{micro}} & 2\,\mu_{\textrm{micro}}+\lambda_{\textrm{micro}} & \star & 0 & 0 & 0\\
\star & \star & \star & 0 & 0 & 0\\
0 & 0 & 0 & \star & 0 & 0\\
0 & 0 & 0 & 0 & \star & 0\\
0 & 0 & 0 & 0 & 0 & \mu_{\textrm{micro}}^{*}
\end{pmatrix},
&\Cte & =\begin{pmatrix}2\me+\le & \le & \star & 0 & 0 & 0\\ \le & 2\me+\le & \star & 0 & 0 & 0\\ \star & \star & \star & 0 & 0 & 0\\ 0 & 0 & 0 & \star & 0 & 0\\ 0 & 0 & 0 & 0 & \star & 0\\ 0 & 0 & 0 & 0 & 0 & \me^{*} \end{pmatrix}, 
\\
\Jth & =\begin{pmatrix}2\eta_{1}+\eta_{3} & \eta_{3} & \star & 0 & 0 & 0\\
\eta_{3} & 2\eta_{1}+\eta_{3} & \star & 0 & 0 & 0\\
\star & \star & \star & 0 & 0 & 0\\
0 & 0 & 0 & \star & 0 & 0\\
0 & 0 & 0 & 0 & \star & 0\\
0 & 0 & 0 & 0 & 0 & \eta_{1}^{*}
\end{pmatrix},
& \Jte & =\begin{pmatrix}2\overline{\eta}_{1}+\overline{\eta}_{3} & \overline{\eta}_{3} & \star & 0 & 0 & 0\\ \overline{\eta}_{3} & 2\overline{\eta}_{1}+\overline{\eta}_{3} & \star & 0 & 0 & 0\\ \star & \star & \star & 0 & 0 & 0\\ 0 & 0 & 0 & \star & 0 & 0\\ 0 & 0 & 0 & 0 & \star & 0\\
0 & 0 & 0 & 0 & 0 & \overline{\eta}_{1}^{*}
\end{pmatrix},
\end{align*}
\begin{align*}
\Ctc & =\begin{pmatrix}\star & 0 & 0\\
0 & \star & 0\\
0 & 0 & 4\mc
\end{pmatrix},  & \Jtc & =\begin{pmatrix}\star & 0 & 0\\
0 & \star & 0\\
0 & 0 & 4\eta_{2}
\end{pmatrix},& \Jtg & =\begin{pmatrix}\star & 0 & 0\\
0 & \star & 0\\
0 & 0 & 4\overline{\eta}_{2}
\end{pmatrix}.
\end{align*}

\section{Dispersive properties of a band-gap metamaterial with tetragonal symmetry \label{sec:bf}}
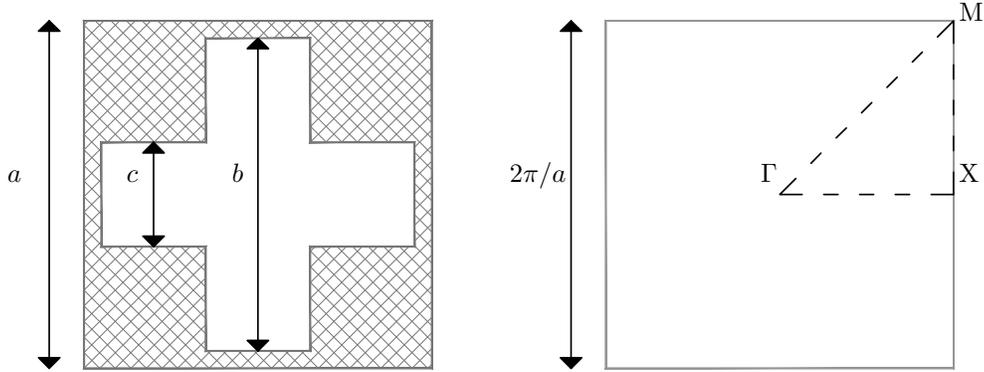
\begin{figure}[h!]
	\begin{centering}
			\includegraphics[height=5.2cm]{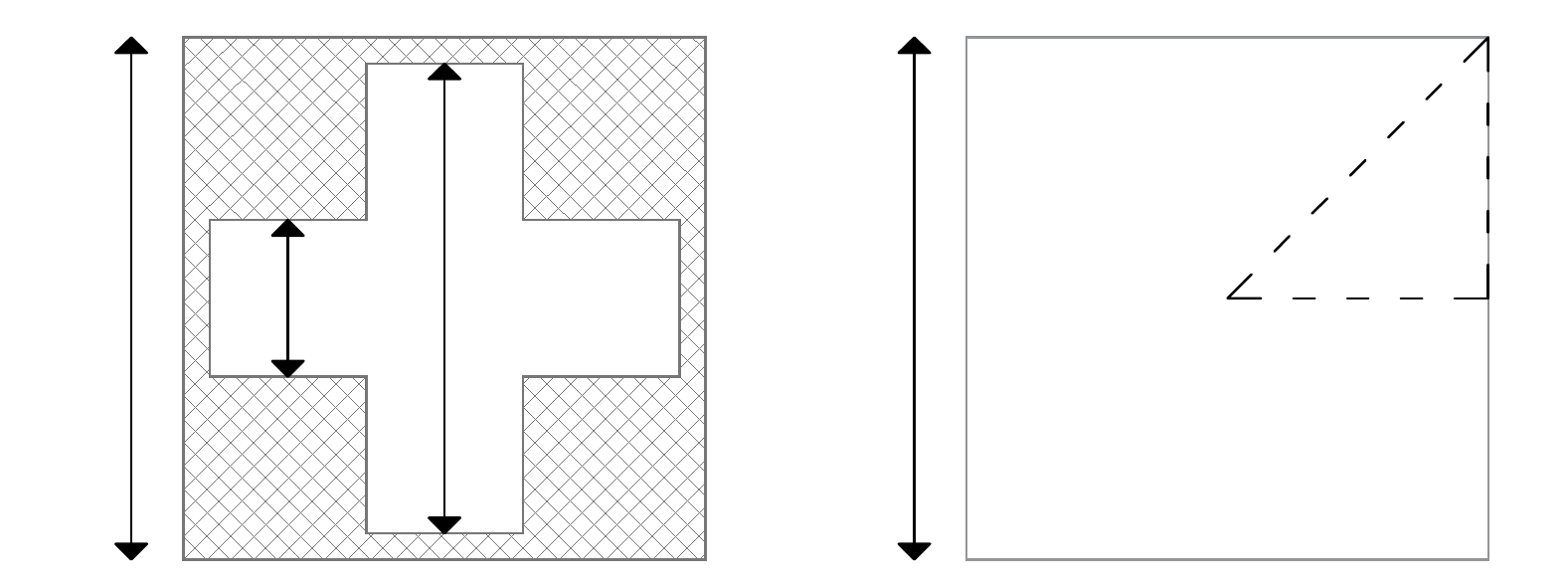} 
		\par\end{centering}
	\caption{\label{fig:Microstructure} Panel (a) is a schematic representation of the unit-cell of the square lattice. In panel (b), the gray square represents the boundaries of the first Brillouin zone. The path connecting the high-symmetry points $\rm \Gamma$, $\rm X$ and $\rm M$,
		on which the dispersion surfaces will be projected  is  also shown (black dashed line).}
	\begin{picture}(0,0)(0,0)
	\put(70,132) {$a$}
	\put(115,132) {$c$}
	\put(155,132) {$b$}
	\put(260,132) {${2\pi}/{a}$}
	\put(355,132) {$\Gamma$}
	\put(430,132) {$\rm X$}
	\put(430,193) {$\rm M$}
	\end{picture}
\end{figure}

In this section, we choose a specific two-dimensional (2D) metamaterial whose dispersive properties and frequency band-gaps have already been investigated in the literature (see \cite{dagostino2018effective} and section \ref{sec:dynamic-determination} of the present article).  The unit-cell of the considered metamaterial is represented in Fig. \ref{fig:Microstructure}(a). 
In Fig. \ref{fig:Microstructure}(a), the shaded areas represent aluminium while the white areas denote cross-shaped holes. In Tab. \ref{parametri microstruttura}, we give the geometric parameters of the unit cell (left), as well as physical parameters for the aluminium (right).
\begin{table}[H]
	\begin{centering}
		\begin{tabular}{ccc}
			$a$ & $b$ & $c$ \tabularnewline[1mm]
			\hline 
			\noalign{\vskip1mm}
			$\left[\mathrm{mm}\right]$ & $\left[\mathrm{mm}\right]$ & $\left[\mathrm{mm}\right]$ \tabularnewline[1mm]
			\hline 
			\hline 
			\noalign{\vskip1mm}
			$1$ & $0.9$ & $0.3$ \tabularnewline[1mm]
		\end{tabular}$\qquad\qquad$%
		\begin{tabular}{ccccc}
			$E$ & $\nu$ & $\mu$ & $\lambda$&$\rho_{\rm Al}$\tabularnewline[1mm]
			\hline 
			\noalign{\vskip1mm}
			$\left[\mathrm{GPa}\right]$ & $-$ & $\left[\mathrm{GPa}\right]$ & $\left[\mathrm{GPa}\right]$&$\left[\mathrm{Kg/m^3}\right]$\tabularnewline[1mm]
			\hline 
			\hline 
			\noalign{\vskip1mm}
			$70$ & $0.33$ & $26.32$ & $51.08$&$2700$  \tabularnewline[1mm]
		\end{tabular}
		\par\end{centering}
	\caption{\label{parametri microstruttura}Geometric parameters of the unit-cell (see also Fig.\,\ref{fig:Microstructure}(a)), elastic parameters and mass density $\rho_{\rm Al}$ of aluminium. For the reader's comfort we give both the values of Young's modulus and Poisson's ratio ($E,\nu$), as well as the Lam$\rm \acute{e}$ parameters ($\lambda,\mu$).}
\end{table}


In the following subsections, we will present the basis to analyse the dynamical behaviour of the unit-cell by means of a Bloch-Floquet analysis (section \ref{sec:bf}), via the relaxed micromorphic model (section \ref{sec:rel}) and  we will show  the resulting dispersion curves in section \ref{sec:dynamic-determination}.

\subsection{Bloch-Floquet  dispersion diagram}
The 2D Bravais lattice with unit-cell in Fig. \ref{fig:Microstructure}(a) has primitive vectors 
\begin{equation}
t_1 = \begin{pmatrix}
1\\0\\0
\end{pmatrix} a,~~~{\rm and}~~~t_2 = \begin{pmatrix}
0\\1\\0
\end{pmatrix} a,
\end{equation} 
and reciprocal lattice primitive vectors 
\begin{equation}\label{eq:primitive_vectors_rec}
G_1 = \frac{2 \pi}{a} \begin{pmatrix}
1\\
0\\
0
\end{pmatrix},~~~{\rm and}~~~G_2 = \frac{2 \pi}{a} \begin{pmatrix}
0\\
1\\
0
\end{pmatrix}.
\end{equation}
The first Brillouin zone for the lattice coincides with the  Wigner-Seitz region of the reciprocal Bravais lattice and is represented in Fig. \ref{fig:Microstructure}(b). We also mark the representative points in the first Brillouin zone 
\begin{equation}
\Gamma=\frac{\pi}{a}\begin{pmatrix}
0\\0\\0
\end{pmatrix},~~~{\rm X}=\frac{\pi}{a}\begin{pmatrix}
1\\0\\0
\end{pmatrix},~~~{\rm M}=\frac{\pi}{a}\begin{pmatrix}
1\\1\\0
\end{pmatrix},
\end{equation}   
which will be used for the representation and discussion of the  dispersive properties of the periodic structure. Those properties are obtained by solving the PDE eigenfrequency  problem governed by  time-harmonic linear plain strain elasticity in the domain $\Omega_0$ represented by the unit cell. 

The Bloch-Floquet boundary conditions for the displacement field in the periodic structure are 
\begin{equation}
u(x+t^{(n)})=e^{i \langle k , t^{(n)} \rangle } u(x),~~~{x}\in\Omega_0, 
\end{equation}
where we have introduced the lattice nodal points 
\begin{equation}
t^{(n)}=n_1t_1+n_2 t_2,~~~n=(n_1,n_2)^{\rm T}\in\mathbb{Z}^2,
\end{equation}
and the Bloch-Floquet wave vector $k=(k_1,k_2,0)^{\rm T}$, with 
\begin{equation}\label{eq:1bz}
(k_1,k_2)^{\rm T}\in[-\frac{\pi}{a},\frac{\pi}{a}]\times[-\frac{\pi}{a},\frac{\pi}{a}].
\end{equation} 
The right-hand-side in Eq \eqref{eq:1bz} is the first Brillouin zone for the square lattice under consideration (see also Fig. \ref{fig:Microstructure}(b) for a schematic representation). We assume that the cross-like holes in Fig \ref{fig:Microstructure}(a) delimit vacuum. Therefore we assign  traction-free boundary conditions in correspondence of the boundary of the cross-like hole in Fig \ref{fig:Microstructure}(a). 
We use COMSOL to solve the Bloch-Floquet eigenvalue problem described above, with spectral parameter $\omega=\omega(k_1,k_2).$

\subsection{Relaxed micromorphic model: identification of the constitutive parameters and the plane wave Ansatz \label{sec:rel}}

In \cite{dagostino2018effective}, an explicit procedure for the \emph{a priori} determination of the material parameters of a specific unit-cell has been proposed. The setup of such procedure is not the main objective of the present paper. Nevertheless, it is worthwhile resuming here such a determination procedure, as it is of crucial importance to understand the interest of using the relaxed micromorphic model with respect to other generalised continuum models available in the literature. In fact, the relaxed micromorphic model features a limited number of constitutive parameters, all with a clear physical meaning, which can be determined \emph{a priori} on the basis of simple static tests. More secifically, $\mathbb{C}_{\rm macro}$ represents the mechanical properties of a very large specimen of the considered metamaterial, $\mathbb{C}_{\rm micro}$ encloses the mechanical properties of smaller specimens, down to the unit cell scale, and $\mathbb{C}_e$ gives the transition between the two scales. As already anticipated, most of the parameters can be determined directly starting from a simulation of the unit-cell. Indeed,
\begin{itemize}
	\item applying Periodic Boundary Conditions (PBCs) to the boundary of the unit-cell, the macroscopic tensor 	$\Csym_{\mathrm{macro}}$ can be determined.  This procedure is equivalent to run a standard mechanical test on a large specimen of the considered metamaterial to obtain its mechanical properties. It is also equivalent to classical homogenisation procedures;
	\item applying Kinematically Uniform (Dirichlet) Boundary Conditions (KUBCs) to the boundary of the unit-cell, the microscopic tensor 	$\Ch$ can be determined (see \cite{dagostino2018effective} for an extensive treatment \color{black}{and, in particular, for a motivation of the choice of KUBC rather than other possible boundary conditions}); \color{black} this procedure is equivalent to running mechanical tests on specimens of the metamaterial whose dimensions are comparable to those of the unit cell;
	\item $\Ce$ is directly computed inverting the harmonic tensor mean \eqref{eq:Relation};
	\item from the cut-off frequencies (obtained via a Bloch-Floquet analysis, see section \ref{sec:dynamic-determination}), we derive the micro-inertiae $\eta_{1},\eta_{3}$ and $\eta_{1}^{*}$	and an explicit relation between the Cosserat couple modulus $\mu_{c}$ and $\eta_{2}$. As a matter of fact, the cut-off frequencies are related to the parameters of the relaxed micromorphic model via the simple formulae
\begin{align}
\omega_{r} & =\sqrt{\frac{\mc}{\eta_{2}}}, & \omega_{s} & =\sqrt{\frac{\me+\mh}{\eta_{1}}}, & \omega_{s}^{*} & =\sqrt{\frac{\me^{*}+\mh^{*}}{\eta_{1}^{*}}}, & \omega_{p} & =\sqrt{\frac{\me+\mh+\le+\lh}{\eta_{1}+\eta_{3}}}.\label{cut-offs}
\end{align}
\end{itemize}
Thus, the only parameters which are still free after the proposed identification procedure are the inertias $\bar{\eta}_1$, $\bar{\eta}_1^*$ and $\bar{\eta}_3$ linked to $\nabla u_{,t}$ and $\eta_{2}$. Those parameters can be used to fine-tune the dispersion curves or some other specific behavior of the propose metamaterial. In Table \ref{tab:Numerical values}, we present the material parameters resulting from the systematic physically-grounded procedure described above (we refer the reader to \cite{dagostino2018effective} for an extensive discussion about this point), as well as the fine tuning of the dispersion curves. The characteristic length $L_c$ is set to be 0.  Nevertheless, a non-vanishing $L_c$ is a crucial point for a description of the non-locality of meta-materials, but the task of its identification is postponed to forthcoming work.

\begin{table}[H]
	\begin{centering}
		\begin{tabular}{ccccc}
			\begin{tabular}{ccc}
				$\le$ & $\mu_{e}$ & $\me^{*}$\tabularnewline[1mm]
				\hline 
				\noalign{\vskip1mm}
				$\left[\textrm{GPa}\right]$ & $\left[\textrm{GPa}\right]$ & $\left[\textrm{GPa}\right]$\tabularnewline[1mm]
				\hline 
				\hline 
				\noalign{\vskip1mm}
				$-\,0.77$ & $17.34$ & $0.67$\tabularnewline[1mm]
			\end{tabular} &   %
			\begin{tabular}{ccc}
				$\lh$ & $\mh$ & $\mh^{*}$\tabularnewline[1mm]
				\hline 
				\noalign{\vskip1mm}
				$\left[\textrm{GPa}\right]$ & $\left[\textrm{GPa}\right]$ & $\left[\textrm{GPa}\right]$\tabularnewline[1mm]
				\hline 
				\hline 
				\noalign{\vskip1mm}
				$5.98$ & $8.93$ & $8.33$\tabularnewline[1mm]
			\end{tabular}&  
			\begin{tabular}{c}
				$\mu_{c}$ \tabularnewline[1mm]
				\hline 
				\noalign{\vskip1mm}
				$\left[\textrm{GPa}\right]$\tabularnewline[1mm]
				\hline 
				\hline 
				\noalign{\vskip1mm}
				$2.2\cdot10^{-2}$\tabularnewline[1mm]
			\end{tabular}& 	\begin{tabular}{c}
			$L_{c}$ \tabularnewline[1mm]
			\hline 
			\noalign{\vskip1mm}
			$\left[\textrm{m}\right]$\tabularnewline[1mm]
			\hline 
			\hline 
			\noalign{\vskip1mm}
			$0$\tabularnewline[1mm]
		\end{tabular}&
			\begin{tabular}{ccc}
				$\lambda_{\mathrm{macro}}$ & $\mu_{\mathrm{macro}}$ & $\mu_{\mathrm{macro}}^{*}$\tabularnewline[1mm]
				\hline 
				\noalign{\vskip1mm}
				$\left[\textrm{GPa}\right]$ & $\left[\textrm{GPa}\right]$ & $\left[\textrm{GPa}\right]$\tabularnewline[1mm]
				\hline 
				\hline 
				\noalign{\vskip1mm}
				$1.74$ & $5.89$ & $0.62$\tabularnewline[1mm]
			\end{tabular}\tabularnewline
		\end{tabular}
		
		\begin{tabular}{ccccccccc}
			$\rho$ & $\eta_{1}$ & $\eta_{2}$ & $\eta_{3}$ & $\eta_{1}^{*}$&$\overline{\eta}_{1}$ & $\overline{\eta}_{2}$ & $\overline{\eta}_{3}$ & $\overline{\eta}_{1}^{*}$\tabularnewline[1mm]
			\hline 
			\noalign{\vskip1mm}
			$\left[\mathrm{kg/m}^{3}\right]$ & $\left[\mathrm{kg/m}\right]$ & $\left[\mathrm{kg/m}\right]$ & $\left[\mathrm{kg/m}\right]$ & $\left[\mathrm{kg/m}\right]$&$\left[\mathrm{kg/m}\right]$ & $\left[\mathrm{kg/m}\right]$ & $\left[\mathrm{kg/m}\right]$ & $\left[\mathrm{kg/m}\right]$\tabularnewline[1mm]
			\hline 
			\hline 
			\noalign{\vskip1mm}
			$1485$ & $9.5\cdot10^{-5}$ & $1\cdot10^{-7}$ & $0.86\cdot10^{-5}$ & $3.27\cdot10^{-5}$&$5\cdot10^{-5}$ & $5\cdot 10^{-5}$  & $8\cdot10^{-5}$& $0$\tabularnewline[1mm]
		\end{tabular}

		\par\end{centering}
	\caption{\label{tab:Numerical values}Summary of the numerical values for the elastic (top) and inertia (bottom) parameters of the tetragonal relaxed micromorphic model in 2D. The macroscopic parameters of the resulting homogenised Cauchy material are also provided. }
\end{table}

The fundamental analytical solutions of the PDEs system (\ref{eq:PDE system}) are given by the plane waves 
\begin{align}
	\u\left(x,t\right) & =\widehat{u}\,e^{i \left( \langle k,  x\rangle -\,\omega t\right)},\qquad P=\widehat{P}\,e^{i \left( \langle k, x \rangle -\,\omega t\right)},\qquad k=(k_1,k_2,0)^{\rm T},\label{eq:wave strucutre-1}
\end{align}
where $x=(x_1,x_2,0)^{\rm T}$, and the plane strain assumption (see section \ref{sec:plane-strain}) implies 
\begin{equation}
\widehat{u}=\begin{pmatrix}
\widehat{u}_1\\\widehat{u}_2\\0
\end{pmatrix},
\qquad\qquad\widehat{P}=\begin{pmatrix}
\widehat{P}_{11}&\widehat{P}_{12}&0\\
\widehat{P}_{21}&\widehat{P}_{22}&0\\
0&0&0
\end{pmatrix}.
\end{equation}
Substituting (\ref{eq:wave strucutre-1}) in (\ref{eq:PDE system}),
the system of PDEs (\ref{eq:PDE system}) turns into an algebraic
problem:
\begin{equation}\label{eq:algebraic_system}
	D\left(k,\omega\right)\, v=0,
\end{equation}
where $D\left(k,\omega\right)$ is a $6\times6$ semi-positive definite matrix  (see \cite{neff2017real}), whose components are functions of the relaxed micromorphic constitutive parameters, of the wave vector $k$ and of frequency $\omega$. The explicit expression of the matrix $D$ is algebraically involved, but easly tractable with standard numerical tools, such as Mathematica or Matlab. For the explicit expression of the matrix $D$, we refer the interested readers to \cite{dagostino2018effective}. The  algebraic system (\ref{eq:algebraic_system}) admits non-trivial solutions $v=(\hat{u}_1,\hat{u}_2,\hat{P}_{11},\hat{P}_{12},\hat{P}_{21},\hat{P}_{22})^{\rm T}$ if and only if  
\begin{equation}\label{eq:disp_eq}
{\rm det}\,D(k,\omega)=0,
\end{equation}
which is referred to as  \textbf{dispersion equation} for the micromorphic continuum. The solutions  $\omega=\omega\left(k_1,k_2\right)$ of Eq. \eqref{eq:disp_eq} provide the \textbf{dispersion relations} for plane wave propagation in a micromorphic continuum. 

We remark that, by choosing a wave vector $k$ and thus a propagation direction, the frequencies $\omega=\omega(k)$ can be obtained as roots of Eq. \eqref{eq:disp_eq}. In the present article we choose to project the dispersion surfaces over the path $k\in$M$\rm \Gamma$XM, by sampling such a path, and iteratively solving Eq. \eqref{eq:disp_eq}.

\subsection{Dispersive properties of the metamaterial and of the equivalent continuum\label{sec:dynamic-determination}}
\begin{figure}[H]
	\begin{centering}
		\includegraphics[width=0.8\linewidth]{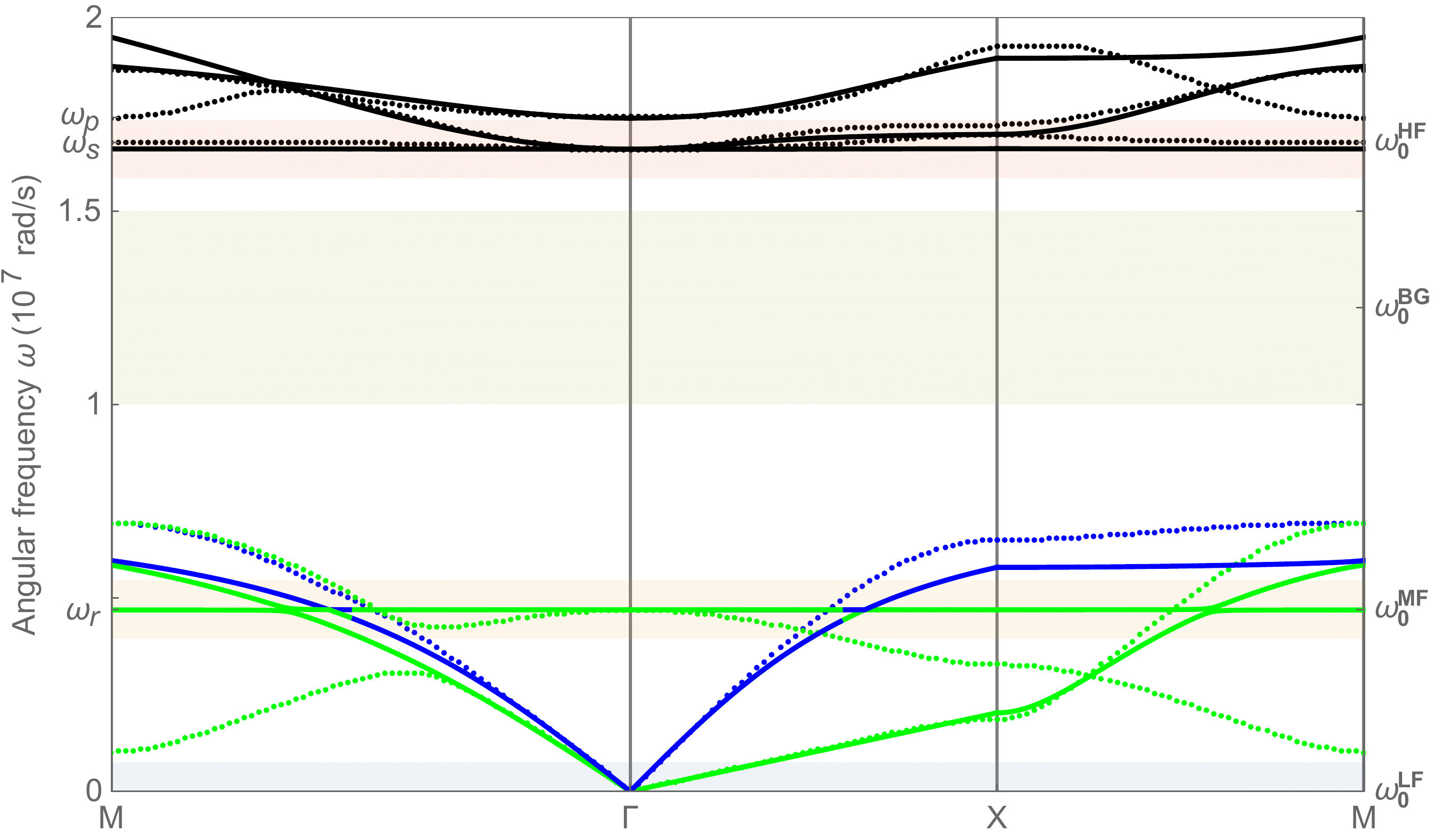} \par\end{centering}
	\caption{\label{fig:dispersion}Comparison of dispersion curves obtained solving Eq. \eqref{eq:disp_eq} for the relaxed micromorphic model (solid lines) and the Bloch-Floquet dispersion diagrams for the micro-structured domain (see section \eqref{sec:bf}) (dotted lines). We project the dispersion curves over the path M$\rm \Gamma$XM in the first Brillouin zone for the periodic structure (see Fig. \ref{fig:Microstructure}(b)). With reference to Tab. \ref{tab:freq-par}, the blue shadow area has thickness $\Delta\omega^{\rm LF}/2$ (the negative part is omitted) and is delimited from below by $\omega_0^{\rm LF}=0$. The vertical thicknesses of the remaining shadows, from bottom to top,  correspond to the values of $\Delta\omega$ and are centred around $\omega_0^{i}$, $i=\{$MF,BG,HF$\}$, respectively (see Tab. \ref{tab:freq-par}). \color{black}{The blue curves refer to pressure dominated modes, while the green curves identify shear-dominated and rotation-dominated modes. Such identification cannot be done at high frequency, where the dispersion branches are represented in black}\color{black}.}
\end{figure}

Fig. \ref{fig:dispersion} shows the dispersion diagrams over the path $k\in$ M$\rm \Gamma$XM in the first Brillouin zone of the considered periodic structure (see the dashed line in Fig. \ref{fig:Microstructure}(b)). The dotted lines in Fig \ref{fig:dispersion} are obtained performing the Bloch-Floquet analysis outlined in section \ref{sec:bf}. The solid lines are the roots $\omega=\omega(k)$ of the relaxed micromorphic dispersion equation \eqref{eq:disp_eq} for $k\in$ M$\rm \Gamma$XM (see section \ref{sec:rel}). For the considered microstructure, we obtain the cut-off frequencies shown in Table \ref{table:cut-offs}.

\begin{table}[H]
	\begin{centering}
		\begin{tabular}{cccc}
			${\omega}_{r}$ & ${\omega}_{s}$ & ${\omega}_{s}^{*}$ & ${\omega}_{p}$\tabularnewline[1mm]
			\hline 
			\noalign{\vskip1mm}
			$\left[\mathrm{rad/s}\right]$ & $\left[\mathrm{rad/s}\right]$ & $\left[\mathrm{rad/s}\right]$ & $\left[\mathrm{rad/s}\right]$\tabularnewline[1mm]
			\hline 
			\hline 
			\noalign{\vskip1mm}
			$0.4\cdot10^{7}$ & $1.68\cdot10^{7}$ & $1.68\cdot10^{7}$ & $1.75\cdot10^{7}$\tabularnewline[1mm]
		\end{tabular}
		\par\end{centering}
	\caption{\label{table:cut-offs}Numerical values of the cut-offs for the considered
		metamaterial.}
\end{table}

It is worthwhile pointing out that the notion of first Brillouin zone for a homogeneous continuum medium such as the relaxed micromorphic model, loses its sense, since a continuum is intrinsically macroscopic and a relation with specific unit cells can only be found ``a posteriori''. On the other hand, dispersion diagrams in periodic media with microstructure are periodic in the reciprocal space identified by $k$, the period being the first Brillouin zone. 

We explicitly remark that the relaxed micromorphic model is able to account for the overall dispersive properties of the metamaterial with unit cell in Fig. \ref{fig:Microstructure}(a) (see Fig. \ref{fig:dispersion}). Some differences can be appreciated especially in the neighbourhood of the boundaries of the first Brillouin zone, corresponding to large values of the Bloch vector. This means that the relaxed micromorphic continuum model is less accurate when considering small wavelengths which become comparable with the size of the unit cell. Even if a loss of precision for small wavelengths is expected when considering continuum models, further generalisations of the relaxed micromorphic model will allow a better precision also for such small wavelengths. Nevertheless, as we will show in the next section, the used homogenised model is precise enough for describing the propagation of pulses in metastructures, also at relatively high frequencies.   

\section{ Description of transient pulse propagation in metastructures via the relaxed micromorphic model \label{sec:pulse-def}}

As pointed out in the introductory section, time-harmonic problems in discrete lattices and micro-structured continua have been extensively studied in the literature. For non-homogeneous elastic problems in infinite micro-structured domains, finite-element time-harmonic solutions heavily rely on the ability to model a perfectly absorbing frame in a finite computational domain. For lattice structures, it is often sufficient to introduce a damping term in the equations of motion for the masses belonging to an exterior  computational frame \cite{colquitt2012dynamic,carta2014dispersion,tallarico2017tilted,tallarico2017edge}. For continua with microstructure, this is not possible and the research is currently ongoing, even if several \emph{ad-hoc} solutions, based on the perfectly matched layer method, have been already proposed \cite{diatta2016scattering}.

\medskip
To overcome this problem, we choose to apply a concentrated pulse in the middle of the considered meta-structure.  We show transient waveforms in the time-domain, before spurious reflections, stemming from the interaction with the boundaries of the computational domains, intervene. This method allows us to avoid the implementation of perfectly absorbing frames, but demands the precise definition of a \emph{space-concentrated} and \emph{time-modulated} prescribed displacement. The remainder of this section is devoted to such a definition.

\subsection{Computational domains}

\begin{figure}[h]
	\begin{centering}
		\includegraphics[width=0.48\linewidth]{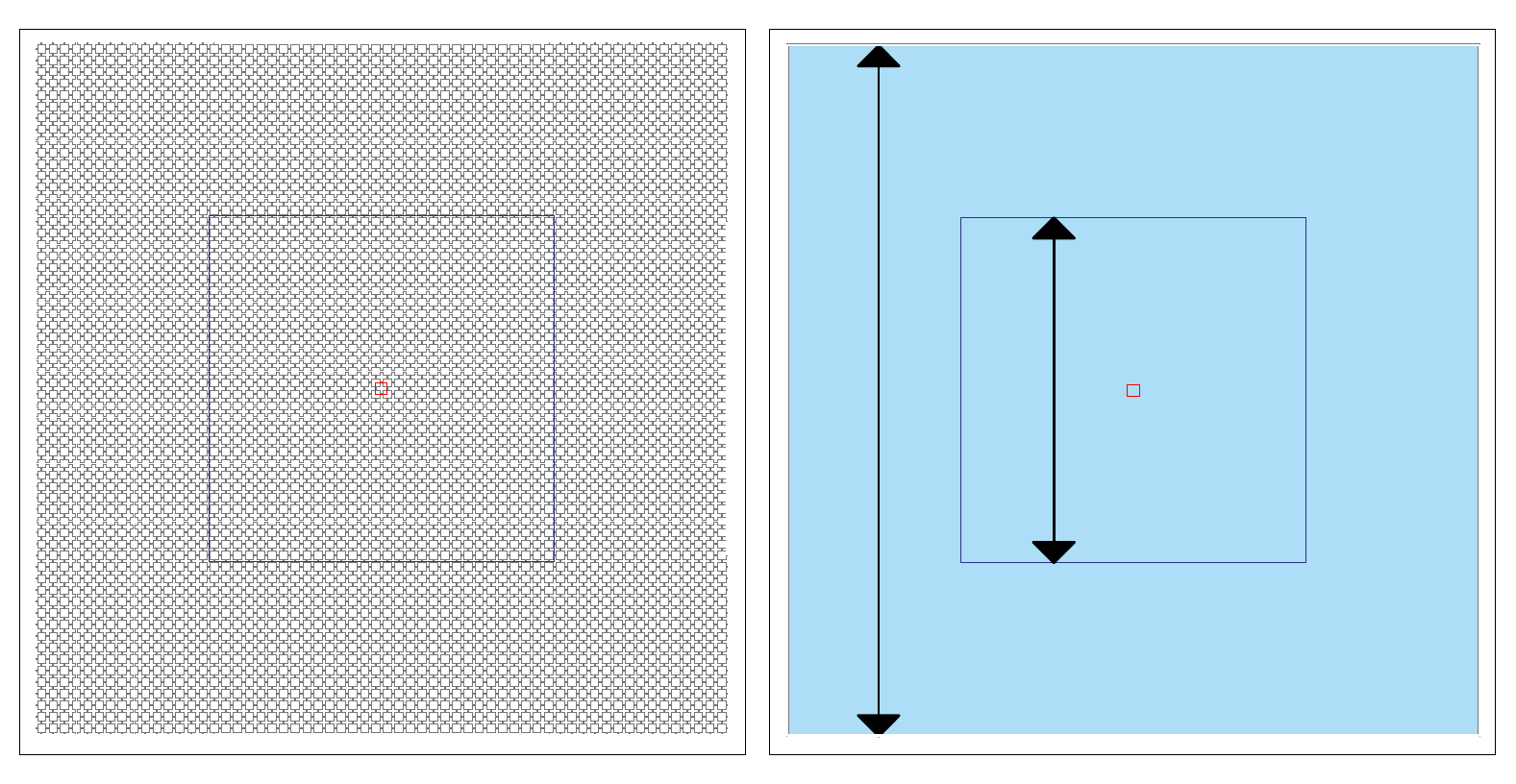} \hspace{-0.325cm}
		\includegraphics[width=0.48\linewidth]{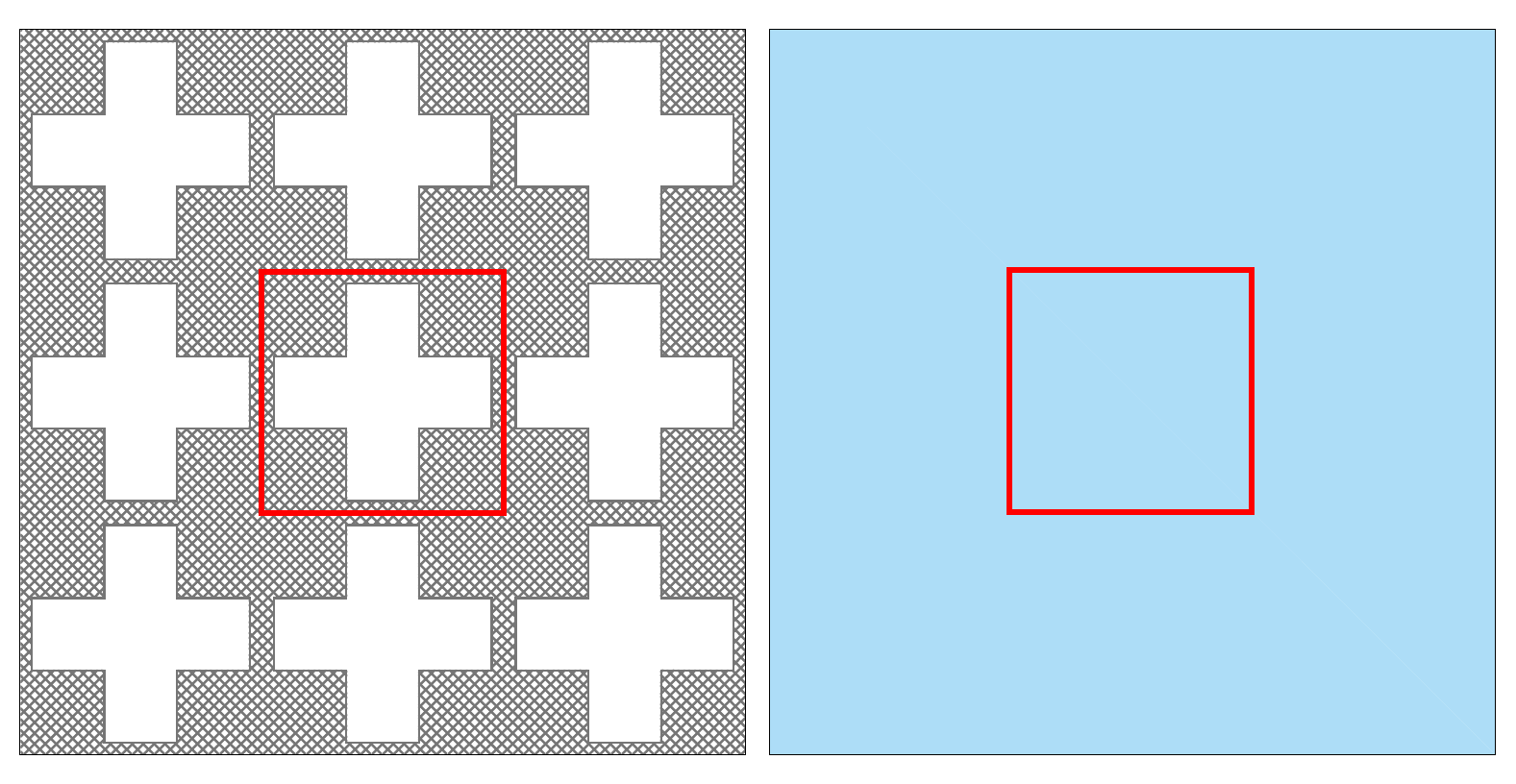} 
		\\
		$(a)$\hspace{0.215\linewidth}$(b)$\hspace{0.215\linewidth}$(c)$\hspace{0.215\linewidth}$(d)$
		\par\end{centering}
	\begin{picture}(0,0)(0,0)
	\put(140,70) {$A$}
	\put(168,70) {$B$}
	\end{picture}
	\caption{\label{fig:fig_geometry} Schematic representations of the computational domains. Panel (a) is a micro-structured domain textured with cross-like holes. Panel (b) is a homogeneous domain modelled as a generalised continuum. The red inner square $\partial \Omega_0$, represents the boundary where the external prescribed displacement is applied. Panels (c) and (d) are magnifications of panels (a) and (b), respectively, comprising the square domain $\partial \Omega_0$. \color{black}{The dimension $A$ is relative to the whole computational domain, while the squares of side B are the windows over which the results are reported (see Figs \ref{fig:pressure-lf}-\ref{fig:shear-hf})}\color{black}.}
\end{figure}
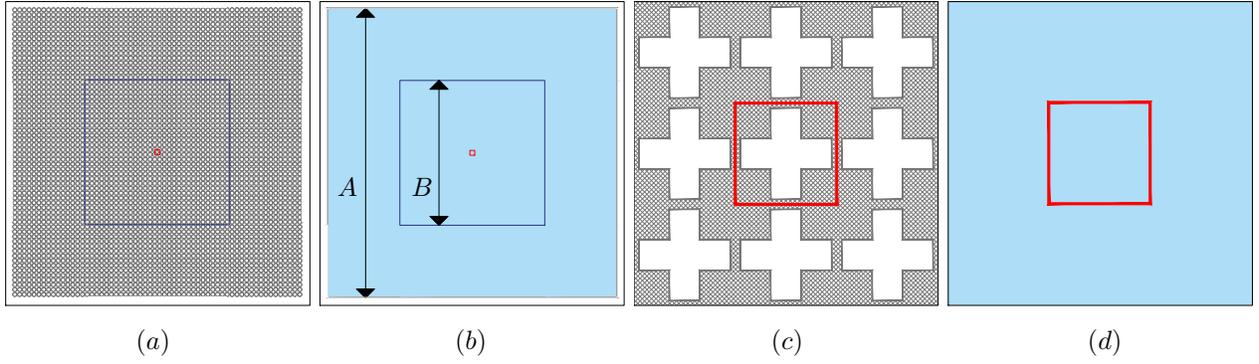

The computational domains implemented in COMSOL Multiphysics are schematically represented in Fig. \ref{fig:fig_geometry}(a) and Fig. \ref{fig:fig_geometry}(b). These domains are squares each of which has side $A=121\,a$ and is centred at the origin of the 2D Cartesian system of coordinates $(x_1,x_2)$. 
Fig. \ref{fig:fig_geometry}(a) is textured with a microstructure, whose magnification is given in Fig. \ref{fig:fig_geometry}(c). Fig. \ref{fig:fig_geometry}(b)  shows a continuous domain in which we solve the system of PDEs \eqref{eq:PDE system} for the relaxed micromorphic model  and the system of PDEs \eqref{eq:PDEs-cmacro} for the equivalent Cauchy  continuum. 

Fig. \ref{fig:fig_geometry}(d) is a  magnification of Fig. \ref{fig:fig_geometry}(b). The solid red squares in Figs  \ref{fig:fig_geometry}(c) and \ref{fig:fig_geometry}(d), here referred to as $\partial \Omega_0$,  denote the boundaries on which the prescribed displacements are applied (see also the boundary conditions in Eqs \eqref{eq:BCs} and \eqref{eq:PDEs-cmacro} for the boundary conditions pertaining the relaxed micromorphic and the equivalent Cauchy models, respectively). 
On the other hand, the micro-structured computational domain in Fig \ref{fig:fig_geometry}(a) is governed by the equations of classical linear elasticity 
\begin{align}\label{eq:PDEs-c}
\rho_{\rm Al}\,u_{,tt}&=\textrm{Div}\left[\mathbb{C}\,\sym\nabla u\right]& \forall x&\in\Omega,\\\nonumber
f:&=\left(\mathbb{C}\,\sym\nabla u\right)\cdot n=t^{\rm ext}(x,t) \qquad \mathrm{or} \qquad u=\varphi(x,t),& \forall x&\in\partial\Omega_0, 
\end{align}
where $\mathbb{C}$ is the stiffness matrix for aluminium, with elastic parameters given in Tab. \ref{parametri microstruttura}. In solving Eq. \eqref{eq:PDEs-c}, traction-free boundary conditions are used at the cross-like holes and at the external boundary $\partial \Omega$\footnote{We explicitly remark that, for the micro-structured solid, the domain $\Omega$ which appears in Eqs \eqref{eq:PDEs-c}  coincides with the grey area in Figs \ref{fig:fig_geometry}(a) and \ref{fig:fig_geometry}(c).}.
In the following we define the time  and space dependence the external prescribed displacement $\varphi$, which appear in Eqs \eqref{eq:BCs}, \eqref{eq:PDEs-cmacro} and \eqref{eq:PDEs-c}.

\subsection{Definition of the concentrated and modulated pulse \label{PulseDef}}
The concentrated and modulated pulse is a prescribed displacement for the systems of PDEs  governing the structures in Fig. \ref{fig:fig_geometry}. We assume that the frequency spectrum of the applied pulses is of the Gaussian-type. The Gaussian function describing the frequency content has mean  $\omega_0$, corresponding to either the acoustic branches,  the band-gap  or  the optical cut-offs of the dispersion diagram in Fig. \ref{fig:dispersion}, see Table \ref{table:cut-offs}. The Gaussian function has standard deviation proportional to  $\Delta\omega$, whose values have been deliberately chosen to be much smaller than the highest cut-off frequency of the structure. The frequency content of the time-dependent part of the prescribed displacement can be represented as the following function
\begin{equation}\label{eq:fourier_transform}
{f}\left(\omega\right)=\frac{2}{\Delta\omega}\,\exp\left[-2\,\left(\frac{\omega-\omega_{0}}{\Delta\omega
		}\right)^{2}\right],\qquad~{\rm with}~\Delta\omega\ll\omega_{p}.
\end{equation}
Since the problem is linear, the choice of a Fourier spectrum like in Eq. \eqref{eq:fourier_transform} guarantees that the structure will respond only to predefined frequency intervals which are  $\Delta \omega$ wide and have $\omega_0$ mean value.

The Fourier anti-transform of the function \eqref{eq:fourier_transform} is 
\begin{equation}\label{eq:pulse-time}
\mathscr{I}\left(t\right)= {\cal F}^{-1}[f(\omega)]=\exp\left[{-i\omega_{0} t-\frac{1}{8}\,\left(\Delta\omega\right)^{2}t^2}\right].
\end{equation}

\begin{figure}[h]
\includegraphics[width=0.47\linewidth]{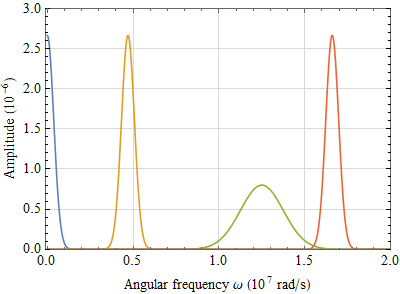}\hfill
\includegraphics[width=0.5\linewidth]{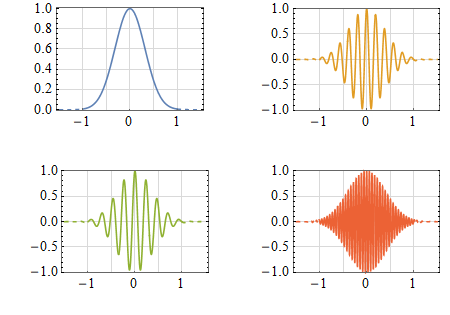}
\begin{picture}(0,0)(0,0)
\put(-188,93) {$t/t_0$}
\put(-55,93) {$t/t_0$}
\put(-185,3) {$t/t_0$}
\put(-55,3) {$t/t_0$}
\put(-455,155) {$(a)$}
\put(-250,165) {$(b)$}
\put(-120,165) {$(c)$}
\put(-250,75) {$(d)$}
\put(-120,75) {$(e)$}
\end{picture}
	\caption{\label{fig:Frequency contents}Frequency contents and time-dependence of four prescribed displacements. Panel (a) shows the time-Fourier transform of the pulses (see Eq. \eqref{eq:fourier_transform}). Blue, orange, green and red curves are the LF, MF, BG, HF regimes, respectively. 
	Panel (b), (c), (d) and (e) are the real parts of the  corresponding  Fourier anti-transforms as a function of time ( \emph{i.e.} ${\rm Re}(\mathscr{I}(t))$, where $\mathscr{I}(t)$ is in Eq. \eqref{eq:pulse-time}). The values of $t_0$ are given in Tab. \ref{tab:freq-par} for each of the four frequency regimes.}
	
\end{figure}
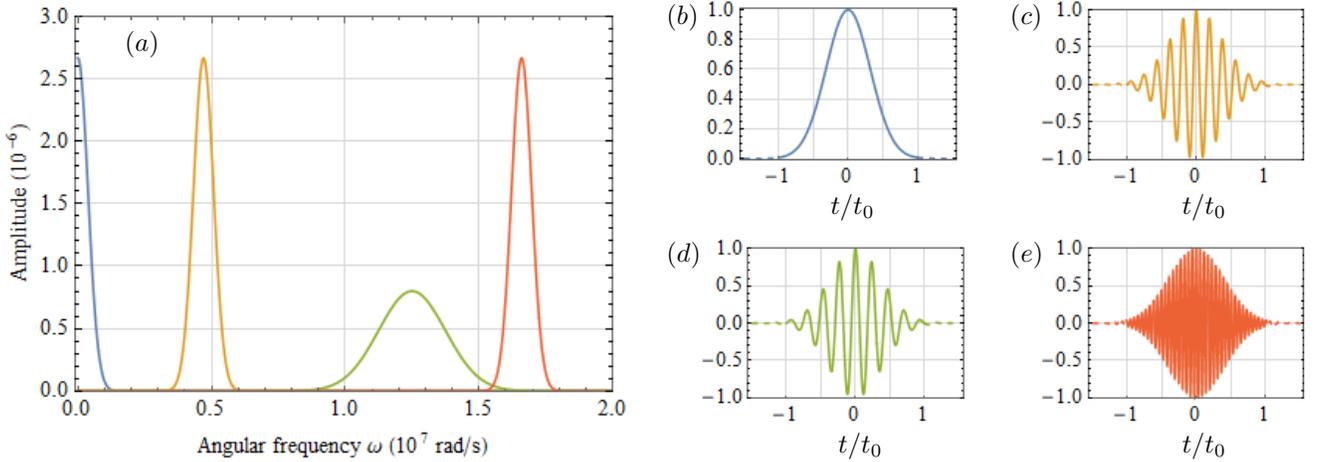

As mentioned earlier, a time-modulated pulse allows us to explore selected frequency regimes in the dispersion diagram. We denote these frequency regimes as low frequency regime (LF), medium frequency regime (MF), band-gap regime (BG) and high frequency (HF) regime. In each of these frequency regimes, the Fourier transform in Eq. \eqref{eq:body-force} has to be considered with different values of $\omega_0$ and $\Delta \omega$. Those values are listed in Tab. \ref{tab:freq-par}. The values of $\omega_0$ are also marked along the $y-$axis of  Fig. \ref{fig:dispersion}.  In Fig. \ref{fig:Frequency contents}(a), we evaluate the functions \eqref{eq:fourier_transform}, in each frequency regime, as a function of frequency: blue, orange, green and red curves correspond to the LF, MF, BG and HF regimes, respectively. Figs  \ref{fig:Frequency contents}(b), \ref{fig:Frequency contents}(c), \ref{fig:Frequency contents}(d) and \ref{fig:Frequency contents}(e) represent the Fourier anti-transform \eqref{eq:pulse-time}  as a function of time, in the LF, MF, BG and HF regimes, respectively. With reference to Fig. \ref{fig:Frequency contents}, in panels (b), (c), (d) and (e) we have introduced a time scale $t_0$.  We choose as initial and final time of the transient pulse the times $t=-t_0$ and $t=t_0$, respectively. The time scale $t_0$ is not arbitrary: with reference to Eq. \eqref{eq:pulse-time}, in order to have $\|\mathscr{I}\left(t\right)\|_{t=\pm t0}\ll1$, \emph{i.e.} a small pulse at the initial and final times, it follows that
\begin{equation}\label{eq:condition-t0}
|t_0|\gg1/\Delta\omega.
\end{equation}

This means that the values of $t_0$ have to be chosen according to the values of $\Delta \omega$, and thus depend on the frequency regime. The values we use for the time scale $t_0$ are reported in the third column of Tab. \ref{tab:freq-par}.

\begin{table}[H]
\centering
\begin{tabular}{l*{6}{c}r}
	& $\omega_0~[10^7~{\rm rad/s}]$  & $\Delta\omega~[10^5{\rm rad/s} ]$  & $t_0~[10^{-6}~{\rm s}]$ \\
	\hline \vspace{-0.3cm}\\
	LF & 0 & 7.5 & 8.38   \\
	MF & 0.47 & 7.5 & 7.02   \\
	BG &  1.25& 25 & 2.14   \\
	HF & 1.66 & 7.5 & 6.72   
\end{tabular}
\caption{\label{tab:freq-par}Parameters used to model the time dependence of the pulse.}
\end{table}

\begin{figure}[H]
	\begin{centering}
		\includegraphics[width=14cm]{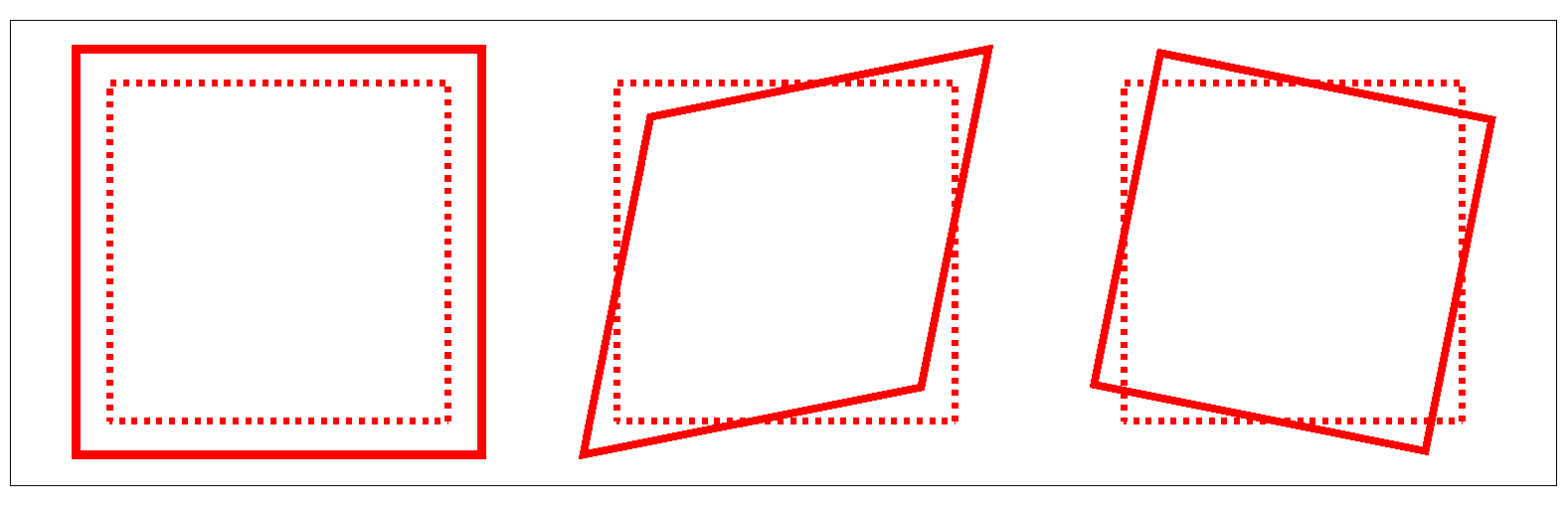} 
		\par\end{centering}
	\caption{\label{fig:fig_geom_pulse} Schematic representations (from left to right) of  pressure-like, shear-like, and rotation-like prescribed displacement. The dotted red square represent the non-deformed square $\partial\Omega_0$ and the solid red contours represent its deformations as prescribed by Eqs \eqref{eq:pres_disp}. The figure is out of scale.}
\end{figure}

The prescribed displacement has  the space and time expression 
\begin{equation}\label{eq:body-force}
\varphi_j\left(x,t\right)= \varphi_0~\mathscr{I}\left(t\right)~S_j(x),\,\,\,x\in\partial \Omega_0,\,\,\,j=\{p,s,r\},
\end{equation}
where the function $\mathscr{I}\left(t\right)$ has been introduced in Eq. \eqref{eq:pulse-time}, the domain $\partial \Omega_0$ corresponds to the red squares in Fig. \ref{fig:Microstructure}, \color{black}{and $\varphi_0=1.4\times10^{-4}~{\rm m}$}\color{black}. The vector functions of the space variable in Eq. \eqref{eq:body-force}  are 
\begin{equation}\label{eq:pres_disp}
S_p(x_1,x_2)=\begin{pmatrix}
x_1/a\\
x_2/a\\
0
\end{pmatrix},\,\,\,S_s(x_1,x_2)=\begin{pmatrix}
x_2/a\\
x_1/a\\
0
\end{pmatrix},\,\,\,S_r(x_1,x_2)=\begin{pmatrix}
x_2/a\\
-x_1/a\\
0
\end{pmatrix},
\end{equation}
where we recall that $a$ is the side of the square unit-cell (see Fig. \ref{fig:Microstructure}(a)). 
The functions of space \eqref{eq:pres_disp} result in a pressure-like  ($p$), shear-like ($s$) or rotation-like ($r$) deformation of the square $\partial \Omega_0$, respectively. A schematic representation of the deformations induced by the functions \eqref{eq:pres_disp} is given in Fig. \ref{fig:fig_geom_pulse}. The dashed red lines in Fig. \ref{fig:fig_geom_pulse} represent the undeformed squares $\partial\Omega_0$ and the solid lines represent, from left to right, pressure-like, shear-like  and rotation-like deformations. 

The advantage of the transient problem with a prescribed displacement as in Eq. \eqref{eq:body-force} is that the anisotropic  propagation or localisation of the waveforms   (both ``fingerprints'' of the non-trivial dispersion in metamaterials) can be studied over finite time intervals, short enough for the solution not to reach the boundary of the computational domain but long enough to allow the signal to develop.

\section{Results and discussion \label{sec:results}}
In this section, we report transient finite element solutions for the following PDE models:
\begin{itemize}
	\item  a fully resolved micro-structured domain, with physical properties as in Table \ref{parametri microstruttura} and geometry  depicted in Fig. \ref{fig:fig_geometry}(a); such medium is governed by the system of PDEs   \eqref{eq:PDEs-c}, with prescribed displacement as in Eq. \eqref{eq:body-force}; 
	\item a relaxed micromorphic model, with physical properties as in Table \ref{tab:Numerical values} and the domain of Fig. \ref{fig:fig_geometry}(b); such enriched continuum is governed by the system of PDEs \eqref{eq:PDE system}, with prescribed displacement as in Eq. \eqref{eq:body-force}; 
	\item an equivalent macroscopic Cauchy continuum governed by $\mathbb{C}_{\rm macro}$ (see Eq. \eqref{eq:Relation} and Table \ref{tab:Numerical values}), in the domain depicted in  Fig. \ref{fig:fig_geometry}(b).	
\end{itemize}

For the relaxed micromorphic model, rather than considering the strong form in Eq. \eqref{eq:PDE system}, we implement the corresponding weak form, readily obtained imposing the first variation of the action functional to be zero (see Eq. \eqref{eq:var-action}), \emph{i.e.}
\begin{equation}\label{eq:var-action}
\delta \left(\int_{0}^{T}{\rm d}t\int_{\Omega}(J-W){\rm d}\Omega\right)=0, 
\end{equation}
where the function $J$ is the kinetic energy density in Eq. \eqref{eq:KineticAniso} and the function $W$ is the potential energy density in Eq. \eqref{eq:EnerRelaxed}. The variation is here intended  with respect to the six independent kinematic fields entering the plane-strain Ansatz in Eq. \eqref{eq:plane-strain}. 

In the following, transient solutions will be given at several times $t$, for the different models mentioned above and a prescribed displacement of the type in Eq. \eqref{eq:body-force}. We consider three different space dependencies (pressure-like, shear-like and rotation like) and the frequency regimes defined in section \ref{PulseDef}, i.e.:
\begin{enumerate}
	\item low frequency (LF);
	\item medium frequency (MF);
	\item band-gap (BG);
	\item high frequency (HF).
\end{enumerate}
Each of the following subsections will be devoted to the analysis of  the transient response of the metastructure in the aforementioned frequency regimes. The time domain is sampled with 10 time steps between the zeros of each signal in Figs \ref{fig:Frequency contents}(b) to \ref{fig:Frequency contents}(e). The results for the modulus of the displacement field are represented on a square frames of side $B=30~a$ (see Figs \ref{fig:fig_geometry}(a) and Figs \ref{fig:fig_geometry}(b)). \color{black}{The fully resolved finite element model solves for 250000 degrees-of-freedom at every time step, whereas only 100000 degrees of freedom are needed to obtain convergent results in the homogenised-Cauchy case and relaxed micromorphic case results.} This reduction is due to the fact that the relaxed micromorphic model is a continuos model, while the full resolution of the real system intrinsically reveals a discrete nature. Such intrinsic simplification related to the use of an enriched continuum modeling will result in a computational time-saving as it will be reported in the results section.\color{black}

 \subsection{Low frequency regime \label{sec:lf}}

In this section, we consider a prescribed displacement as in Eq. \eqref{eq:body-force} in the low frequency regime (see the nomenclature introduced in section \ref{sec:pulse-def}). Specifically, the frequency content is centred around zero (see blue curve in Fig. \ref{fig:Frequency contents}(a)), which results in a Gaussian pulse in the time domain (see Fig. \ref{fig:Frequency contents}(b)).

\begin{figure}[H]
	\begin{centering}
		 \vspace{-0.25cm}
		\begin{tabular}{c|c|c|c|c}
			\footnotesize LF&$t=-\frac{1}{2} t_0$ & $t=0$ &$t=\frac{1}{2}t_0$& $t=t_0$ \vspace{-0.55cm}\\  	\includegraphics[width=0.7cm,trim={0.6cm 0.5cm 11cm 0.5cm},clip]{./Cross-Modes}  	
			&&&\\ \hline  \vspace{-0.25cm}&&&\\
			\rotatebox{90}{micro-structured} 
			&\includegraphics[width=2.5cm]{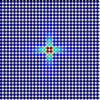} & 	\includegraphics[width=2.5cm]{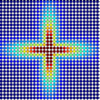} & 	\includegraphics[width=2.5cm]{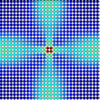} &
			\includegraphics[width=2.5cm]{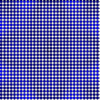} \\ \hline \vspace{-0.25cm}&&&
			\\
			\rotatebox{90}{\hspace{0.05cm}Relaxed Micro}&	\includegraphics[width=2.5cm]{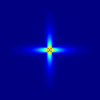} & 	\includegraphics[width=2.5cm]{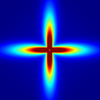} & 	\includegraphics[width=2.5cm]{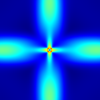} &
			\includegraphics[width=2.5cm]{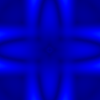} \\\hline\vspace{-0.25cm}&&&
			\\
			\rotatebox{90}{\hspace{0.7cm}Cauchy}    &  \includegraphics[width=2.5cm]{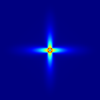} & 	\includegraphics[width=2.5cm]{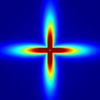} & 	\includegraphics[width=2.5cm]{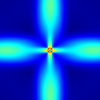} &
			\includegraphics[width=2.5cm]{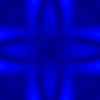} 
			\tabularnewline
		\end{tabular}
		\par\end{centering}
	\caption{\label{fig:pressure-lf} \textbf{Pressure-like} pulse propagation in the \textbf{low frequency} regime. The color map represents \color{black}{modulus of the displacement, with dark blue corresponding to zero  displacement and dark red corresponding to the  $|u_{\rm max}|=3\times10^{-5}~{\rm m}$}. This color scale is adopted in the remainder of the article.~\color{black}}
\end{figure}

The first case considered is the  pressure-like prescribed displacement, obtained inserting the first of Eqs \eqref{eq:pres_disp} in Eq. \eqref{eq:body-force}. In Fig. \eqref{fig:pressure-lf}, we present snapshots of the resulting transient displacement for the micro-structured domain, the relaxed micromorphic continuum and a continuum governed by Cauchy elasticity, respectively. The directions of propagation are given by the properties of the micro-structure and, in particular, the wave propagates mainly on the directions of the beams in the micro-structure (0 and 90 degrees). As expected in the low frequency regime, the anisotropy is very well described by both the relaxed micromorphic model and the macroscopic Cauchy model. The columns correspond to different times, namely $t/t_0=\{-1/2,0,1/2,1\}$ ($t_0$ is the LF value in Tab. \ref{tab:freq-par}). The modulus of the displacement in each of the four different columns match very well. This proves that the propagation speed is very well described in the LF regime for pressure-like excitations.

\begin{figure}[H]
	\begin{centering}
		 \vspace{-0.25cm}
		\begin{tabular}{c|c|c|c|c}
			\footnotesize LF&$t=-\frac{1}{2} t_0$ & $t=0$ &$t=\frac{1}{2}t_0$& $t=t_0$  \vspace{-0.55cm}\\  	
			\includegraphics[width=0.7cm,trim={5.8cm 0.5cm 5.6cm 0.5cm},clip]{./Cross-Modes}  
			&&&\\ \hline  \vspace{-0.25cm}&&&
			\\ 
			\rotatebox{90}{micro-structured} 
			&
			\includegraphics[width=2.5cm]{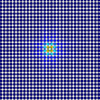} & 	\includegraphics[width=2.5cm]{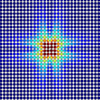} & 	\includegraphics[width=2.5cm]{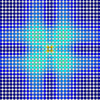} &
			\includegraphics[width=2.5cm]{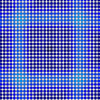}
			\\ \hline \vspace{-0.25cm}&&&
			\\
			\rotatebox{90}{\hspace{0.05cm}Relaxed Micro}
			&
			\includegraphics[width=2.5cm]{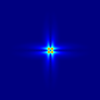} & 	\includegraphics[width=2.5cm]{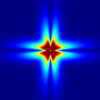} & 	\includegraphics[width=2.5cm]{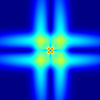} &
			\includegraphics[width=2.5cm]{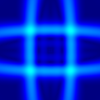}\\\hline\vspace{-0.25cm}&&&
			\\
			\rotatebox{90}{\hspace{0.7cm}Cauchy}    & 
			\includegraphics[width=2.5cm]{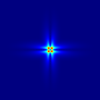} & 	\includegraphics[width=2.5cm]{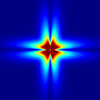} & 	\includegraphics[width=2.5cm]{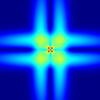} &
			\includegraphics[width=2.5cm]{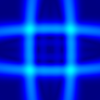} 
			\tabularnewline
		\end{tabular}
		\par\end{centering}
	\caption{\label{fig:shear-lf} \textbf{Shear-like} pulse propagation in the \textbf{low frequency} regime.
	}
\end{figure}

\begin{figure}[H]
	\begin{centering}
		 \vspace{-0.25cm}
		\begin{tabular}{c|c|c|c|c}
	\footnotesize LF&$t=-\frac{1}{2} t_0$ & $t=0$ &$t=\frac{1}{2}t_0$& $t=t_0$  \vspace{-0.55cm}\\  	
	\includegraphics[width=0.7cm,trim={11cm 0.5cm 0.6cm 0.5cm},clip]{./Cross-Modes}  
	&&&\\ \hline  \vspace{-0.25cm}&&&
	\\ 
	\rotatebox{90}{micro-structured} 
	&
			\includegraphics[width=2.5cm]{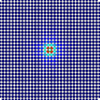} & 	\includegraphics[width=2.5cm]{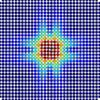} & 	\includegraphics[width=2.5cm]{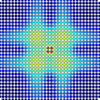} &
			\includegraphics[width=2.5cm]{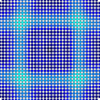}
			\\ \hline \vspace{-0.25cm}&&&
			\\
			\rotatebox{90}{\hspace{0.05cm}Relaxed Micro}
			&
			\includegraphics[width=2.5cm]{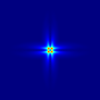} & 	\includegraphics[width=2.5cm]{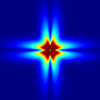} & 	\includegraphics[width=2.5cm]{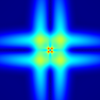} &
			\includegraphics[width=2.5cm]{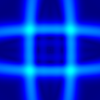}  \\\hline\vspace{-0.25cm}&&&
			\\
			\rotatebox{90}{\hspace{0.7cm}Cauchy}    & 
			\includegraphics[width=2.5cm]{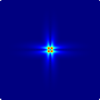} & 	\includegraphics[width=2.5cm]{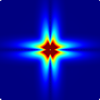} & 	\includegraphics[width=2.5cm]{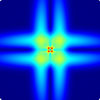} &
			\includegraphics[width=2.5cm]{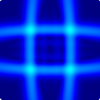} 
			\tabularnewline
		\end{tabular}
		\par\end{centering}
	\caption{\label{fig:rotation-lf}\textbf{Rotation-like}  pulse propagation in the \textbf{low frequency} regime.
	}
\end{figure}

Similar conclusions apply to Figs \ref{fig:shear-lf} and \ref{fig:rotation-lf}, where,  shear-like and rotation-like pulse propagations, respectively, are shown. The direction of propagation is, once again, given by the micro-structure, leading to a relevant energy channeling at $\pm45$ degrees, in both cases. Furthermore, we observe that there is no substantial difference between the computations in Fig. \ref{fig:shear-lf} pertaining to a shear-like pulse  and in Fig. \ref{fig:rotation-lf} pertaining to a rotation-like pulse. In our opinion, the close similarity of the represented waveforms is due to the very low rotational stiffness of the unit cell leading to a shear-like propagation also for rotation-like pulses. The rotational mode, that is described by the skew-symmetric part of $\nablau$ and $P$, has a very low stiffness also in the chosen material parameters of the relaxed micromorphic model ($\mu_c\simeq \mu_{\text{macro}}/200$). The substantial equivalence between shear-like and rotation-like excitation is preserved in all the frequency regimes. Therefore, we will refrain from showing the rotation-like excitations in the remainder of the paper.

Overall, the presented results show that, in the low frequency regime, there is no significant difference between the three examined models. It also confirms a standard low frequency homogenisation result according to which the low frequency dispersion, and the corresponding PDEs, for a metamaterial with tetragonal symmetry are well captured by a macroscopic Cauchy continuum with the same symmetry. Moreover, in \cite{barbagallo2017transparent} it was proven that $\mathbb{C}_{\rm macro}$ corresponds to a macroscopic limit case for the relaxed micromorphic model and, in  \cite{dagostino2018effective}, it has already been shown that the low frequency dispersion of the considered meta-structure is captured by the equations of classical linear elasticity with effective tensor $\mathbb{C}_{\rm macro}$ in Eq. \eqref{eq:Relation}. More specifically, it is shown in \cite{dagostino2018effective} that, at low frequencies, the dispersion curves of the equivalent Cauchy continuum governed by $\mathbb{C}_{\rm macro}$ provide an excellent approximation for the low frequency acoustic branches of the micromorphic continuum.  By means of Fig. \eqref{fig:pressure-lf},  \eqref{fig:shear-lf} and  \eqref{fig:rotation-lf}, we provide a further illustration of this result in the context of low-frequency transient solutions of PDEs. 

\newpage
\subsection{Medium frequency regime}
In this section, we investigate the pulse propagation in the medium frequency regime. This is done by prescribing a displacement as in Eq. \eqref{eq:pres_disp}, with a time-dependence as  in Fig. \ref{fig:Frequency contents}(c). The time-dependence has a spectral content as the orange curve in Fig. \ref{fig:Frequency contents}(a). The spectral width ($\Delta\omega_0^{\text{MF}}$) and centre ($\omega_0^{\text{MF}}$) of this signal is also shown in Fig. \ref{fig:dispersion} by the orange shadow area  (see MF-row in Tab. \ref{tab:freq-par}).
\begin{figure}[H]
	\begin{centering}
		 \vspace{-0.25cm}
		\begin{tabular}{c|c|c|c|c}
		 \footnotesize MF&$t=-\frac{1}{2} t_0$ & $t=0$ &$t=\frac{1}{2}t_0$& $t=t_0$  \vspace{-0.55cm}\\  	\includegraphics[width=0.7cm,trim={0.6cm 0.5cm 11cm 0.5cm},clip]{./Cross-Modes}  	
			&&&\\ \hline  \vspace{-0.25cm}&&&\\
			\rotatebox{90}{micro-structured} &
			\includegraphics[width=2.5cm]{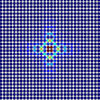} & 	\includegraphics[width=2.5cm]{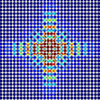} & 	\includegraphics[width=2.5cm]{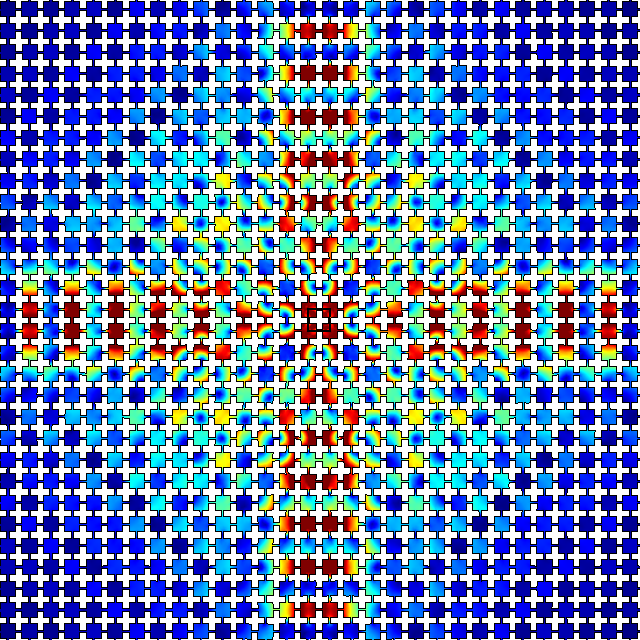} &
			\includegraphics[width=2.5cm]{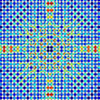}\\ \hline \vspace{-0.25cm}&&&
			\\
			\rotatebox{90}{\hspace{0.05cm}Relaxed Micro}&
			\includegraphics[width=2.5cm]{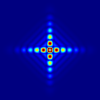} & 	\includegraphics[width=2.5cm]{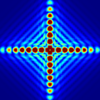} & 	\includegraphics[width=2.5cm]{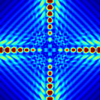} &
			\includegraphics[width=2.5cm]{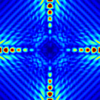} \\\hline\vspace{-0.25cm}&&&
			\\
			\rotatebox{90}{\hspace{0.7cm}Cauchy}  &
			\includegraphics[width=2.5cm]{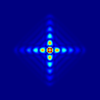} & 	\includegraphics[width=2.5cm]{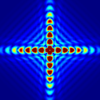} & 	\includegraphics[width=2.5cm]{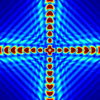} &
			\includegraphics[width=2.5cm]{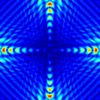} 
			\tabularnewline
		\end{tabular}
		\par\end{centering}
	\caption{\label{fig:pressure-mf}\textbf{Pressure-like} pulse propagation in the \textbf{medium frequency} regime.}
\end{figure}
\begin{figure}[H]
	\begin{centering}
		 \vspace{-0.25cm}
	\begin{tabular}{c|c|c|c|c}
		\footnotesize MF&$t=-\frac{1}{2} t_0$ & $t=0$ &$t=\frac{1}{2}t_0$& $t=t_0$  \vspace{-0.55cm}\\  	
		\includegraphics[width=0.7cm,trim={5.8cm 0.5cm 5.6cm 0.5cm},clip]{./Cross-Modes}  
		&&&\\ \hline  \vspace{-0.25cm}&&&\\
		\rotatebox{90}{micro-structured} &
			\includegraphics[width=2.5cm]{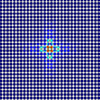} & 	\includegraphics[width=2.5cm]{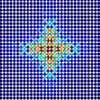} & 	\includegraphics[width=2.5cm]{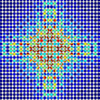} &
			\includegraphics[width=2.5cm]{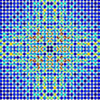}\\ \hline \vspace{-0.25cm}&&&
			\\
			\rotatebox{90}{\hspace{0.05cm}Relaxed Micro}&
			\includegraphics[width=2.5cm]{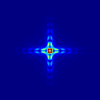} & 	\includegraphics[width=2.5cm]{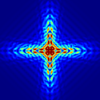} & 	\includegraphics[width=2.5cm]{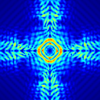} &
			\includegraphics[width=2.5cm]{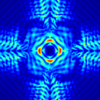} \\\hline\vspace{-0.25cm}&&&
			\\
			\rotatebox{90}{\hspace{0.7cm}Cauchy}  &
			\includegraphics[width=2.5cm]{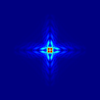} & 	\includegraphics[width=2.5cm]{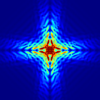} & 	\includegraphics[width=2.5cm]{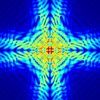} &
			\includegraphics[width=2.5cm]{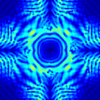} 
			\tabularnewline
		\end{tabular}
		\par\end{centering}
	\caption{\label{{fig:shear-mf}}\textbf{Shear-like} pulse propagation in the \textbf{medium frequency} regime.}
\end{figure}
In Fig. \ref{fig:pressure-mf}, we report snapshots for the transient displacement  as a result of a pressure-like displacement in Eq. \eqref{eq:body-force} ($j=p$ in Eq. \eqref{eq:pres_disp}), using the first of Eqs \eqref{eq:pres_disp}. The rows correspond to the micro-structured domain, relaxed micromorphic domain and Cauchy-elastic domain governed by $\mathbb{C}_{\rm macro}$, respectively. Different columns correspond to different times $t/t_0=\{-1/2,0,1/2,1\}$. With reference to Figs \ref{fig:pressure-mf}, we observe that the anisotropic transient waveforms in the micro-structured domain, are well captured by those pertaining to the relaxed micromorphic model. Moreover, the Cauchy continuum governed by $\mathbb{C}_{\rm macro}$  still represents an acceptable approximation for the waveforms in the micro-structured domain, as it can be appreciated by direct comparison of  the first and the third rows of Figs \ref{fig:pressure-mf}. As shown in Fig. \ref{fig:dispersion}, the frequency mean value of the considered pulse ($\omega_0^{\rm MF}$) as well as its spectral width $(\Delta\omega_0^{\rm MF})$, are located in correspondence of the acoustic branches of the dispersion diagram, in a regime where such branches have a non-linear behaviour as a function of the wave vector $k$. It is well known that, when the dispersion curves deviate from a linear behaviour, the group velocity is not a constant along the direction prescribed by $k$, thus depending on its modulus $|k|=2\pi/\lambda_0$, with $\lambda_0$ being the wavelength.  In this case the medium is said to be \emph{dispersive}. It is also very well known that the group velocity for a Cauchy infinite medium is constant along the direction prescribed by $k$, which translates into the fact that the dispersion curves as a function of $k=|k|\hat{k}$ - where $\hat{k}$ is a fixed unit direction - are simply straight lines. Hence, a Cauchy medium is referred to as \emph{dispersionless}. Even if the relaxed micromorphic model is able to account for the aforementioned dispersion, we observe that this fact does not improve the descriptive power of the relaxed micromorphic model in the medium frequency regime with respect to the Cauchy one. At this stage,  we would like to recall that a relaxed micromorphic model could be considered to be equivalent to a second gradient model for relatively low frequencies, where the dispersion diagram comprises acoustic branches only.  On the other hand, second gradient models are not able to describe the behaviour of the metastructure at higher frequencies as instead the relaxed micromorphic can do. It is then doubtful how second gradient theories could give additional interesting information concerning wave propagation in real metastructures at frequencies higher than the Cauchy-like regime, and at least for the considered symmetry class.

\subsection{Band-gap regime}
The advantage of using the relaxed micromorphic model can be directly appreciated in the band-gap regime for the periodic structure (see Fig. \ref{fig:dispersion}). This section is devoted to such a frequency regime. To investigate the band-gap regime, we design a pulse with frequency content as represented by the green curve in Fig. \eqref{fig:Frequency contents}(a). The corresponding time dependence of the pulse is reported in Fig.  \ref{fig:Frequency contents}(d) and it is used in Eq. \eqref{eq:body-force} as a prescribed displacement over the square $\partial \Omega_0$. 
\begin{figure}[H]
	\begin{centering}
		 \vspace{-0.25cm}
	\begin{tabular}{c|c|c|c|c}
		\footnotesize BG&$t=-\frac{1}{2} t_0$ & $t=0$ &$t=\frac{1}{2}t_0$& $t=t_0$  \vspace{-0.55cm}\\  	\includegraphics[width=0.7cm,trim={0.6cm 0.5cm 11cm 0.5cm},clip]{./Cross-Modes}  	
		&&&\\ \hline  \vspace{-0.25cm}&&&\\
		\rotatebox{90}{micro-structured} &
			\includegraphics[width=2.5cm]{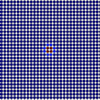} & 	\includegraphics[width=2.5cm]{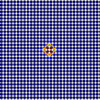} & 	\includegraphics[width=2.5cm]{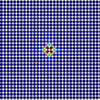} &
			\includegraphics[width=2.5cm]{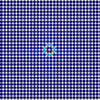} \\ \hline \vspace{-0.25cm}&&&
			\\
			\rotatebox{90}{\hspace{0.05cm}Relaxed Micro}&
			\includegraphics[width=2.5cm]{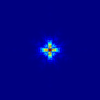} & 	\includegraphics[width=2.5cm]{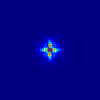} & 	\includegraphics[width=2.5cm]{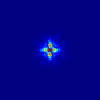} &
			\includegraphics[width=2.5cm]{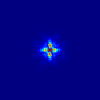} \\\hline\vspace{-0.25cm}&&&
			\\
			\rotatebox{90}{\hspace{0.7cm}Cauchy}  &
			\includegraphics[width=2.5cm]{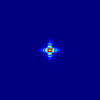} & 	\includegraphics[width=2.5cm]{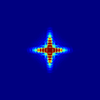} & 	\includegraphics[width=2.5cm]{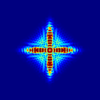} &
			\includegraphics[width=2.5cm]{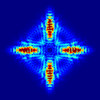} 
			\tabularnewline
		\end{tabular}
		\par\end{centering}
	\caption{\label{fig:pressure-bg}\textbf{Pressure-like} pulse propagation in the \textbf{band-gap} regime.}
\end{figure}
Fig. \ref{fig:pressure-bg} shows the propagation of a pressure-like pulse ($j=p$ in Eq. \eqref{eq:body-force}), in the first row a micro-structured domain, in the second row a continuous domain governed by the relaxed micromorphic model and in the last row  a linear elastic continuous domain governed by $\mathbb{C}_{\rm macro}$. We observe that the pulse is localised throughout all the considered times, in both the  micro-structured medium (panels (a)) and in the relaxed micromorphic medium (panels (b)). This agrees with the fact that $\omega_0^{\text{BG}}$ is in the band gap for both the micro-structured medium and for the equivalent relaxed micromorphic continuum (see Fig. \ref{fig:dispersion}). This is not the case for the panels (c), where the same pulse can propagate in the computational domain. This is because a Cauchy medium, although able to capture anisotropy as shown earlier, is not able  to reproduce band-gaps, leading to pulse propagation at every frequency.    Fig. \ref{fig:shear-bg} shows the propagation of a shear-like pulse ($j=s$ in Eq. \eqref{eq:body-force}). Similar considerations, already done for Fig. \ref{fig:pressure-bg}, apply for the shear-like pulse-localisation phenomena (Panels (a) and (b)) which are witnessed in Fig. \eqref{fig:shear-bg}.  
\begin{figure}[H]
	\begin{centering}
		 \vspace{-0.25cm}
	\begin{tabular}{c|c|c|c|c}
		\footnotesize BG&$t=-\frac{1}{2} t_0$ & $t=0$ &$t=\frac{1}{2}t_0$& $t=t_0$  \vspace{-0.55cm}\\  	
		\includegraphics[width=0.7cm,trim={5.8cm 0.5cm 5.6cm 0.5cm},clip]{./Cross-Modes}  
		&&&\\ \hline  \vspace{-0.25cm}&&&\\
		\rotatebox{90}{micro-structured} &
			\includegraphics[width=2.5cm]{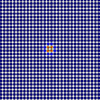} & 	\includegraphics[width=2.5cm]{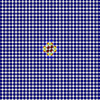} & 	\includegraphics[width=2.5cm]{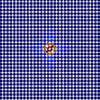} &
			\includegraphics[width=2.5cm]{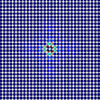} \\ \hline \vspace{-0.25cm}&&&
			\\
			\rotatebox{90}{\hspace{0.05cm}Relaxed Micro}&
			\includegraphics[width=2.5cm]{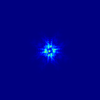} & 	\includegraphics[width=2.5cm]{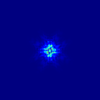} & 	\includegraphics[width=2.5cm]{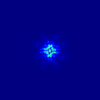} &
			\includegraphics[width=2.5cm]{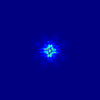} \\\hline\vspace{-0.25cm}&&&
			\\
			\rotatebox{90}{\hspace{0.7cm}Cauchy}  &
			\includegraphics[width=2.5cm]{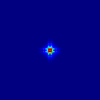} & 	\includegraphics[width=2.5cm]{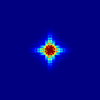} & 	\includegraphics[width=2.5cm]{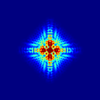} &
			\includegraphics[width=2.5cm]{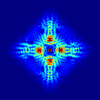} 
			\tabularnewline
		\end{tabular}
		\par\end{centering}
	\caption{\label{fig:shear-bg}\textbf{Shear-like} pulse propagation in the \textbf{band-gap} regime.}
\end{figure}
\subsection{High frequency regime}
In this section, we conclude our analysis by presenting pulse propagation results in the high-frequency regime for the micro-structured domain and for the corresponding relaxed micromorphic continuum (see red shadow in Fig. \ref{fig:dispersion}). This is done by considering a frequency content for the pulse as the red curve in Fig. \eqref{fig:Frequency contents}(a). The corresponding time-dependence, shown in Fig. \eqref{fig:Frequency contents}(e), is assigned to the prescribed displacement in Eq. \eqref{eq:body-force}.
\begin{figure}[H]
	\begin{centering}
		 \vspace{-0.25cm}
	\begin{tabular}{c|c|c|c|c}
		\footnotesize HF&$t=-\frac{1}{2} t_0$ & $t=0$ &$t=\frac{1}{2}t_0$& $t=t_0$  \vspace{-0.55cm}\\  	\includegraphics[width=0.7cm,trim={0.6cm 0.5cm 11cm 0.5cm},clip]{./Cross-Modes}  	
		&&&\\ \hline  \vspace{-0.25cm}&&&\\
		\rotatebox{90}{micro-structured} &
			\includegraphics[width=2.5cm]{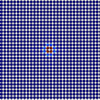} & 	\includegraphics[width=2.5cm]{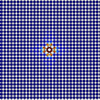} & 	\includegraphics[width=2.5cm]{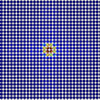} &
			\includegraphics[width=2.5cm]{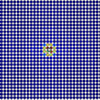} \\ \hline \vspace{-0.25cm}&&&
			\\
			\rotatebox{90}{\hspace{0.05cm}Relaxed Micro}&
			\includegraphics[width=2.5cm]{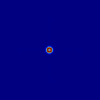} & 	\includegraphics[width=2.5cm]{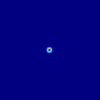} & 	\includegraphics[width=2.5cm]{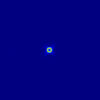} &
			\includegraphics[width=2.5cm]{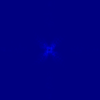}  \\\hline\vspace{-0.25cm}&&&
			\\
			\rotatebox{90}{\hspace{0.7cm}Cauchy}    &  \includegraphics[width=2.5cm]{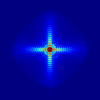} & 	\includegraphics[width=2.5cm]{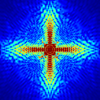} & 	\includegraphics[width=2.5cm]{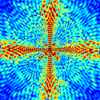} &
			\includegraphics[width=2.5cm]{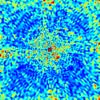} 
			\tabularnewline
		\end{tabular}
		\par\end{centering}
	\caption{\label{fig:pressure-hf}\textbf{Pressure-like} pulse propagation in the \textbf{high frequency} regime. }
\end{figure}
Similarly to what we have done in the previous frequency regimes, we here focus on a pressure-like prescribed displacement (Fig. \ref{fig:pressure-hf}) and on a shear-like prescribed displacement (Fig. \ref{fig:shear-hf}). As suggested by the zero group velocity in the dispersion diagram, we expect almost complete localisation. The localisation is observed for both the pressure-like and shear-like pulses, in both the micro-structured domain and in the relaxed micromorphic continuum. On the other hand, in the Cauchy continuum governed by $\mathbb{C}_{\rm macro}$, the localisation is completely lost since the dispersion is linear in all the investigated  frequency regimes, and hence propagation of waves is always expected. 
\begin{figure}[H]
	\begin{centering}
		 \vspace{-0.25cm}
	\begin{tabular}{c|c|c|c|c}
		\footnotesize HF&$t=-\frac{1}{2} t_0$ & $t=0$ &$t=\frac{1}{2}t_0$& $t=t_0$  \vspace{-0.55cm}\\  	
		\includegraphics[width=0.7cm,trim={5.8cm 0.5cm 5.6cm 0.5cm},clip]{./Cross-Modes}  
		&&&\\ \hline  \vspace{-0.25cm}&&&\\
		\rotatebox{90}{micro-structured} &
			\includegraphics[width=2.5cm]{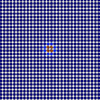} & 	\includegraphics[width=2.5cm]{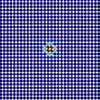} & 	\includegraphics[width=2.5cm]{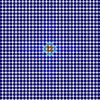} &
			\includegraphics[width=2.5cm]{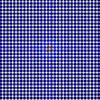}\\ \hline \vspace{-0.25cm}&&&
			\\
			\rotatebox{90}{\hspace{0.05cm}Relaxed Micro}&
			\includegraphics[width=2.5cm]{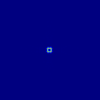} & 	\includegraphics[width=2.5cm]{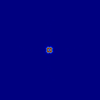} & 	\includegraphics[width=2.5cm]{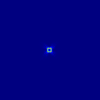} &
			\includegraphics[width=2.5cm]{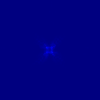}  \\\hline\vspace{-0.25cm}&&&
			\\
			\rotatebox{90}{\hspace{0.7cm}Cauchy}  &
			\includegraphics[width=2.5cm]{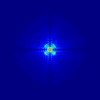} & 	\includegraphics[width=2.5cm]{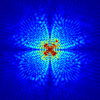} & 	\includegraphics[width=2.5cm]{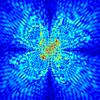} &
			\includegraphics[width=2.5cm]{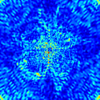} 
			\tabularnewline
		\end{tabular}
		\par\end{centering}
	\caption{\label{fig:shear-hf}\textbf{Shear-like} pulse propagation in the \textbf{high frequency} regime.}
\end{figure}
\section{Conclusion and further perspectives \label{sec:conclusions}}
In this article, we investigate the pertinence of using the relaxed micromorphic model to study wave propagation in band-gap anisotropic metastructures. To this aim, we compare transient waveforms corresponding to a pulse prescribed to the centre of metastructures governed by several equations of motion. The considered equations of motion are (i) a micro-structured domain governed by linear elasticity; (ii) the relaxed micromorphic model; and (iii) an equivalent Cauchy continuum having the same macroscopic stiffness of the analogous relaxed micromorphic continuum. \color{black}{In this article, we show that the relaxed micromorphic model qualitatively captures the behaviour of transient waveforms arising in metastructures, including localised waveforms in the band-gap and in the high frequency regime. The comparison has been performed using several types of prescribed displacements, namely pressure-like, shear-like and rotation-like deformations. We observe that the waveforms arising from rotation-like prescribed displacements - in every frequency regime - are identical to the waveforms obtained via the corresponding shear-like prescribed displacement. Such  substantial equivalence between shear-like and rotation-like waveforms is remarked for the simple metastructure studied in the present paper, and would deserve a deeper investigation to check if it is also possible in more complex situations.} \color{black}

We clearly show that the Cauchy equivalent continuum is able to describe the pulse propagation through the micro-structured domain only in the low frequency regime. The Cauchy equivalent continuum rapidly looses its predictive power when higher frequencies are considered. On the other hand, we show that the relaxed micromorphic model is able to account for the overall behaviour of the metastructure for all the considered frequency regimes, including the band-gap and higher frequencies. This predictive ability is peculiar of the relaxed micromorphic model and is unrivalled by any other generalised continuum models. For example, for the considered metastructure, the so called ``second gradient'' models would not be able to add anything more to its description when compared to the equivalent Cauchy model.  \color{black}{Second gradient models are based on a kinematical framework which is identical to Cauchy classical theory (only macroscopic displacements are provided) and the only difference with the latter consists in a modified constitutive law which also contains second gradients of displacement, instead of only first gradients. In the context of plane-wave propagation and dispersive properties of the generalised continuum, this means that only acoustic modes can be described by a second gradient medium and the only difference with Cauchy media would be the description of non-linear dispersion for high wavenumbers. Failure to capture optical branches and finite frequency band-gaps, represents an intrinsic limitation of second gradient models when applied to description of metamaterials with complex microstructure, even at relatively low frequencies.}\color{black}

The results presented in this paper strongly encourage the promotion of the relaxed micromorphic model to engineering science. Indeed, the following challenges can be identified and will be treated in forthcoming works:
\begin{itemize}
\item study the behaviour of more ``geometrically complex'' metastructures \color{black}(\emph{e.g.} morphologically complex 2D or simplified 3D)\color{black}; this will allow to demonstrate impressive time savings in computations considering that, even for the relatively  simple metastructure presented here, we obtain a considerable reduction in computational time, from 53 minutes in the micro-structured case down to 18 minutes for the relaxed micromorphic case;
\item study the effect of the characteristic length scale $L_c$ on description of the dynamic response of metastructures. \color{black}{ We will show in a forthcoming paper that the characteristic length plays a non-negligible role when considering diffractive phenomena at an interface embedded in a relaxed micromorphic model. Some non-negligible effects of Lc are also expected for very high frequency and will be investigated in further works.}\color{black}
\item study transient wave propagation in metastructures whose unit cell belongs to different symmetry classes; this will demonstrate the unprecedented advantage of using the relaxed micromorphic model rather than other enriched continua. In fact, due to its simplicity compared to existing generalised continua - the relaxed micromorphic model features 4th order tensors rather than 6th order tensors, as in the second gradient model -  the relaxed micromorphic model is able to add tangible improvements to the classical and (infinitely simpler) Cauchy description.
\end{itemize}

\section*{Acknowledgements}
Angela Madeo and Domenico Tallarico acknowledge funding from the French Research Agency ANR, ``METASMART''  (ANR-17CE08-0006). Angela Madeo acknowledges support from IDEXLYON in the framework of the ``Programme Investissement d'Avenir'' ANR-16-IDEX-0005. All the authors acknowledge funding from the ``Région Auvergne-Rhône-Alpes'' for the ``SCUSI'' project for international mobility France/Germany.

{\footnotesize{}\bibliographystyle{plain}
\bibliography{references_pulse}
}{\footnotesize\par}
\end{document}